%% file: elsarticle-3.tex
\documentclass[preprint,12pt]{elsarticle}

\usepackage{amssymb}
\usepackage[cmex10]{amsmath}
\usepackage{amsmath,amsfonts,amssymb,amsthm,dsfont,stmaryrd}
\usepackage{hyperref}

\newtheorem{observation}{Observation}[section]
\newtheorem{theorem}{Theorem}[section]
\newtheorem{proposition}{Proposition}[section]
\newtheorem{lemma}{Lemma}[section]
\newtheorem{corollary}{Corollary}[section]

\newtheorem{definition}{Definition}[section]
\newtheorem{remark}{Remark}[section]
\newtheorem{example}{Example}[section]

\def\s{s}%{{s\I}}
\def\cl{\mathrm{gr}}

\def\R{\mathbb{R}}
\def\Nor{\mathcal{N}\!}

\def\MLE{\mathrm{MLE}}

\def\I{\mathrm{I}}

\def\m{\mathrm{m}}

\def\wcss{\mathbf{ss}}

\def\F{\mathcal{F}}
\def\G{\mathcal{G}}

\def\V{\mathcal{V}}

\def\S{\mathcal{G}_{(\cdot \I)}}%{\mathcal{S}}

\def\e{\varepsilon}
\def\d{\delta}

\def\y{\mathrm{Y}}

\def\1{\mathds{1}}

\def\for{\mbox{  for }}

\def\supp{\mathrm{supp}}
\def\card{\mathrm{card}}

\def\det{\mathrm{det}}
\def\tr{\mathrm{tr}}

\def\sh{\mathrm{sh}}

\numberwithin{equation}{section}

\usepackage{algorithmic}
\usepackage{algorithm}

\newcommand\ced[1]{H^{\times}_d\!\big(\mu \,\| \,  #1 \big)}
\newcommand\ce[1]{H^{\times}\!\big(\mu\|   #1 \big)}

\DeclareMathOperator*{\oti}{\uplus}
\DeclareMathOperator*{\ocup}{\cup}
\def\otis{\uplus}

\usepackage[tight,footnotesize]{subfigure}

\journal{Pattern Recognition}

\begin{document}

\begin{frontmatter}

%% Title, authors and addresses

%% use the tnoteref command within \title for footnotes;
%% use the tnotetext command for the associated footnote;
%% use the fnref command within \author or \address for footnotes;
%% use the fntext command for the associated footnote;
%% use the corref command within \author for corresponding author footnotes;
%% use the cortext command for the associated footnote;
%% use the ead command for the email address,
%% and the form \ead[url] for the home page:
%%
%% \title{Title\tnoteref{label1}}
%% \tnotetext[label1]{}
%% \author{Name\corref{cor1}\fnref{label2}}
%% \ead{email address}
%% \ead[url]{home page}
%% \fntext[label2]{}
%% \cortext[cor1]{}
%% \address{Address\fnref{label3}}
%% \fntext[label3]{}

\title{Cross-Entropy Clustering}

%----------Author 1
\author{J. Tabor}
\address{Faculty of Mathematics and Computer Science, 
Jagiellonian University, 
\L ojasiewicza 6, 
30-348 Krak\'ow, 
Poland}
\ead{jacek.tabor@ii.uj.edu.pl}

%% use optional labels to link authors explicitly to addresses:
%% \author[label1,label2]{<author name>}
%% \address[label1]{<address>}
%% \address[label2]{<address>}
%----------Author 2
\author{P. Spurek}
\address{
Faculty of Mathematics and Computer Science, 
Jagiellonian University, 
\L ojasiewicza 6, 
30-348 Krak\'ow, 
Poland}
\ead{przemyslaw.spurek@ii.uj.edu.pl}

%\author{}

%\address{}

\input{abstract.tex}

\begin{keyword}
%% keywords here, in the form: keyword \sep keyword
clustering \sep cross-entropy \sep memory compression
%% MSC codes here, in the form: \MSC code \sep code
%% or \MSC[2008] code \sep code (2000 is the default)

\end{keyword}

\end{frontmatter}

%%
%% Start line numbering here if you want
%%
% \linenumbers

%% main text
\input{section/intro.tex}

\input{section/cross.tex}
\input{section/many.tex}
\input{section/cluster.tex}

%% The Appendices part is started with the command \appendix;
%% appendix sections are then done as normal sections
%% \appendix

%% \section{}
%% \label{}

%% References
%%
%% Following citation commands can be used in the body text:
%% Usage of \cite is as follows:
%%   \cite{key}          ==>>  [#]
%%   \cite[chap. 2]{key} ==>>  [#, chap. 2]
%%   \citet{key}         ==>>  Author [#]

%% References with bibTeX database:

\bibliographystyle{model1-num-names}
\bibliography{bib/paper}

%% Authors are advised to submit their bibtex database files. They are
%% requested to list a bibtex style file in the manuscript if they do
%% not want to use model1-num-names.bst.

%% References without bibTeX database:

% \begin{thebibliography}{00}

%% \bibitem must have the following form:
%%   \bibitem{key}...
%%

% \bibitem{}

% \end{thebibliography}

\end{document}

%% file: abstract.tex
\begin{abstract}
We build a general and highly applicable clustering theory, which we call cross-entropy clustering (shortly
CEC) which joins advantages of classical k-means (easy implementation
and speed) with those of EM (affine invariance and ability to adapt
to clusters of desired shapes). Moreover, contrary to k-means and EM, {\em CEC
finds the optimal number of clusters by automatically removing groups which carry no information}. 

Although CEC, similarly like EM, can be build on an arbitrary family of densities, in the most important case of Gaussian CEC the division into clusters is affine invariant, while the numerical complexity is comparable to that of k-means.
%\item the clustering has the tendency to divide the data into ellipsoid-type shapes.
%
%We study also with particular attention clustering based on the 
%Spherical Gaussian densities
%and that of Gaussian densities with covariance $s \I$. In the letter case we show that with $s$ converging
%to zero we obtain the classical k-means clustering.
\end{abstract}

%% file: section/intro.tex
%%%%%%%%%%%%%%%%%%%%%%%%%%%%%%%%%%%%
\section{Introduction}

\subsection{Motivation}

\begin{figure}[!t] \centering
	\subfigure[Mouse-like set.]{\label{fig:vor_1_a}\fbox{\includegraphics[width=0.23\textwidth]{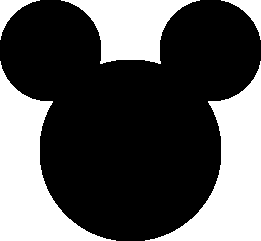}}}
	\subfigure[k-means with $k=3$.]{\label{fig:vor_2_a}\fbox{\includegraphics[width=0.23\textwidth]{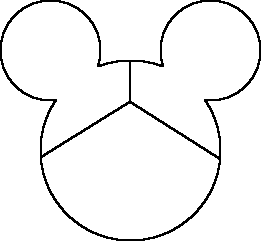}}}
	\subfigure[k-means with $k=10$.]{\label{fig:vor_2_b}\fbox{\includegraphics[width=0.23\textwidth]{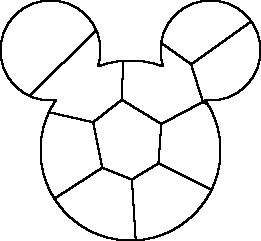}}}
	\subfigure[Spherical CEC.]{\label{fig:vor_3_a}\fbox{\includegraphics[width=0.23\textwidth]{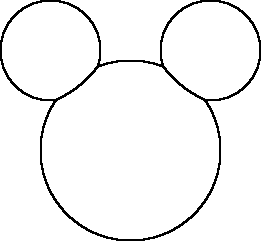}}}			
	\caption{Clustering of the uniform density on mouse-like set (Fig. \ref{fig:vor_1_a}) by standard k-means algorithm with $k=3$ (Fig. \ref{fig:vor_2_a}) and $k=10$ (Fig. \ref{fig:vor_2_b}) compared with Spherical CEC (Fig. \ref{fig:vor_3_a}) with initially $10$ clusters (finished with $3$).}
	\label{fig:mouse} 
\end{figure}

As is well-known, clustering plays a basic role in many parts
of data engineering, pattern recognition and image analysis
\cite{Clu, Clus, jain1999, jain2010,xu2009clustering}. Thus it is not surprising that there are many methods of data clustering, many of which however inherit
the deficiencies of the first method 
called k-means \cite{bock2007, bock2008}.
%\begin{minipage}{\textwidth}
Since k-means has the tendency to divide the data into
spherical shaped clusters of similar sizes, it is not affine invariant and does deal well with clusters of various sizes. This causes the so-called mouse-effect, see
Figure \ref{fig:vor_2_a}.
Moreover, it does not find the right number of clusters, 
see \ref{fig:vor_2_b}, and consequently to
apply it we usually need to use additional tools like gap statistics
\cite{gap,mirkin2011choosing}. 
%\end{minipage}
Since k-means has so many disadvantages, one can
ask why it is so popular. One of the possible answers lies in the fact
that k-means is simple to implement and very fast comparing
to more advanced clustering methods like EM and classification EM
\cite{EM2, EM3}.

So let us now discuss EM, the other end approach to clustering. It
is based on family of densities $\F$ which convex combination 
we allow to estimate the density of the data-set we study. By modifying $\F$ we can adapt our method to the search of clusters of various types
\cite{Ce-Go}. The disadvantages follow from the fact that EM is relatively slow and not well-adapted to dealing with large 
data-sets\footnote{The disadvantages of common clustering methods are excellently summarized in the third paragraph of \cite{estivill2000fast}:
"[...] The weaknesses of k-MEANS result in poor quality clustering, and
thus, more statistically sophisticated alternatives have been proposed.
[...] While these alternatives offer more
statistical accuracy, robustness and less bias, they trade this for substantially more computational
requirements and more detailed prior knowledge \cite{massa1999new}."
}. 
Let us also add that EM, analogously as k-means, does not find
the right number of clusters.

In our paper we construct a 
general cross-entropy clustering (CEC) theory which
simultaneously joins, and even overcomes, the clustering advantages of classical k-means and EM.
The aim of this paper is to study the theoretical
background of cross-entropy clustering. Due to its
length we decided to illustrate it only on basic examples\footnote{For the sample application of CEC in classification and recognition of elliptic shapes we refer the reader to \cite{Ta-Mi}}.

\subsection{Main idea}

We based CEC on the observation that
it is often profitable to use various compression algorithms
specialized in different data types. We apply this observation in reverse, 
namely {\em we group/cluster those data together which are compressed by 
one algorithm from the preselected set of compressing 
algorithms\footnote{We identify a coding/compressing algorithm with a subdensity, see the next section for detailed explanations.}.}
In development of this idea we were influenced 
by the classical Shannon Entropy Theory \cite{Co-Th, Ka, Ku,Sh} 
and Minimum Description Length Principle \cite{MDLP, Gr}. In particular we were strongly inspired by the application of MDLP to image
segmentation given in \cite{Yi_Ma,Yi_Ma2}.

From theoretical point of view our basic idea lies in 
applying cross-entropy to many ``compressing''
densities. Its greatest advantage is the automatic reduction of unnecessary clusters: contrary to the case of classical k-means or EM, there is a memory cost of using each cluster. Consequently from cross-entropy clustering point of view it is in many cases profitable to decrease the number of used clusters.

\begin{example}
\begin{figure}[!t]
\centering
%	\subfigure[Triangle-like set.]{\label{fig:sq_2_a}
%\fbox{\includegraphics[height=0.2\textheight]{figure_1/troj}}}
%	\subfigure[3 clusters.]{\label{fig:sq_3_a}
%\fbox{\includegraphics[height=0.2\textheight]{figure_1/troj-klasy}}}
	\subfigure[Four gaussians.]{\label{fig:gauss_1}
\fbox{\includegraphics[height=0.15\textheight]{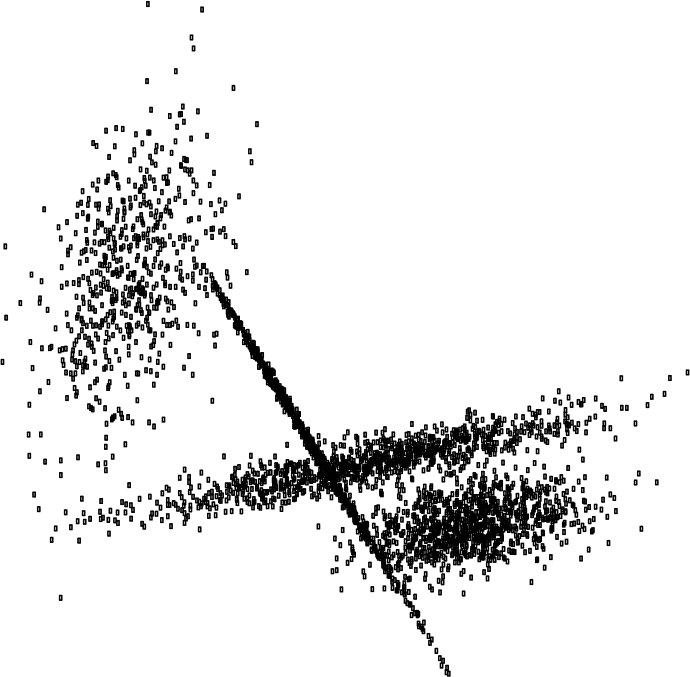}}}
	\subfigure[4 clusters]{\label{fig:gauss_1}
\fbox{\includegraphics[height=0.15\textheight]{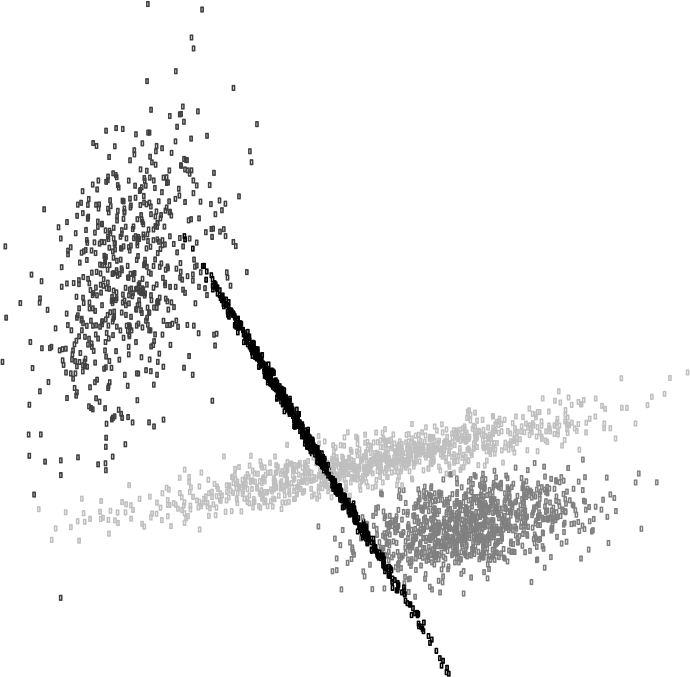}}}		
	\subfigure[Bear-like set.]{\label{fig:bear_1}
\fbox{\includegraphics[height=0.15\textheight]{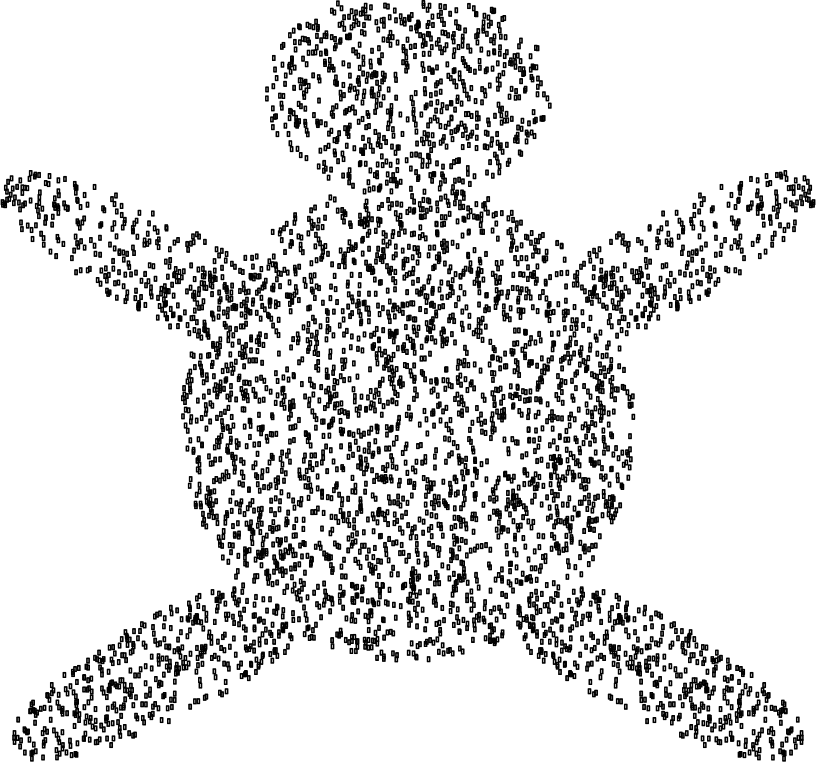}}}
	\subfigure[6 clusters.]{\label{fig:bear_2}
\fbox{\includegraphics[height=0.15\textheight]{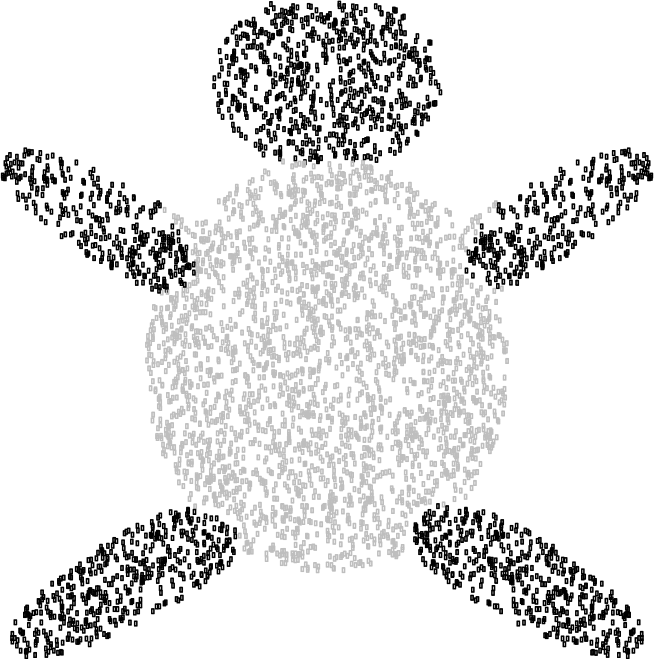}}}
\caption{Gaussian CEC starting from $10$ initial clusters.} 
\label{fig:gauss} 
\end{figure}
To visualize our theory let us
look at the results of Gaussian CEC given in Figure \ref{fig:gauss}. In both cases we started with $k=10$ initial 
randomly chosen clusters which were reduced automatically by the algorithm.
\end{example}

%[[[\textbf{dorzucic wykres z MLE i z mojej metody dla gestosci
% $2x$ dla $x \in [0,1]$!}]]]

In practical implementations our approach can viewed as a generalized and
``modified'' version of the classical k-means clustering 
%\cite{Clu, Clus, bock2007, bock2008} 
As a consequence the complexity of the CEC is usually that of k-means and
one can easily adapt most ideas used in various versions of k-means to CEC.

Since CEC is in many aspects influenced by EM 
%\cite{Mc-Kr, EM2, EM3} 
let us briefly  summarize the main similarities and differences.
Suppose that we are given a continuous probability measure $\mu$ (which represents our data) with density $f_\mu$ and fixed densities $f_1,\ldots,f_k$ by combination of which we want to approximate $f_\mu$. 
{\em \begin{itemize} 
\item The basic goal of EM is to find probabilities $p_1,\ldots,p_k$ such that the approximation	 \begin{equation} \label{ee}
f_\mu \approx p_1f_1+\ldots+p_kf_k
\end{equation}
is optimal. 
\item In CEC we search for partition of $\R^N$
into (possibly empty) pairwise disjoint sets $U_1,\ldots,U_k$ and probabilities $p_1,\ldots,p_k$  such that the approximation 
\begin{equation} \label{eee}
f_\mu \approx p_1f_1|_{U_1} \cup \ldots \cup p_kf_k|_{U_k}
\end{equation}
is optimal.
\end{itemize}}
Observe that as a result of CEC we naturally obtain the partition of the space into sets $(U_i)_{i=1}^k$. Another crucial consequence of the formula \eqref{eee} is that contrary to the earlier approaches based on MLE we { approximate $f_\mu$ not by a density, as is the case in \eqref{ee}, but subdensity\footnote{By subdensity we understand a measurable nonnegative function with integral not greater then one.}}.

\subsection{Contents of the paper}

For the convenience of the reader we now briefly summarize the
contents of the article. In the following section
we discuss (mostly) known results concerning entropy which
we will need in our study. In particular we identify the subdensities
with coding/compressing algorithms. In the third section we provide a detailed motivation and explanation of our basic idea, which allows to interpret the cross-entropy for the case of many ``coding densities''.
More precisely, given a partition $U_1,\ldots,U_k$ of $\R^N$
and subdensity families $\F_1,\ldots,\F_k$,
we introduce the subdensity family
%\begin{equation} \label{jee}
$\oti\limits_{i=1}^k (\F_i|U_i)$,
%\end{equation}
which consists of those acceptable codings in which elements of $U_i$ are compressed by a fixed element from $\F_i$. We also show how to apply classical
Lloyds and Hartigan approaches to cross-entropy minimizations.

The last section contains applications of our theory to clustering.
We first consider a general idea of {\em cross-entropy $\F$-clustering}, 
which aim is to find a $\mu$-partition $(U_i)_{i=1}^k$ minimizing
$$
\ce{\oti_{i=1}^k (\F|{U_i})}.
$$
This allows to investigate the {\em $\F$-divergence of the $\mu$-partition
$(U_i)_{i=1}^k$}
$$
d_{\mu}(\F;(U_i)_{i=1}^k):=
\ce{\F}-\ce{\oti_{i=1}^k (\F|{U_i})}
$$
which measures the validity of the clustering $(U_i)_{i=1}^k$. 

Next we proceed to the study of clustering with respect to various Gaussian subfamilies.
First we investigate the most important case of Gaussian CEC and show that it reduces to the search for the partition $(U_i)_{i=1}^k$ of the given data-set $U$ for which the value of
$$
\sum_{i=1}^k p(U_i) \cdot [-\ln(p(U_i))+\frac{1}{2} \ln \det(\Sigma_{U_i})]
$$
is minimal, where $p(V)=\card(V)/\card(U)$ and $\Sigma_V$ denotes
the covariance matrix of the set $V$. It occurs that the Gaussian
clustering is affine invariant. 

Then we study clustering based on the 
Spherical Gaussians, that is those with covariance proportional to identity.
Comparing Spherical CEC to classical k-means we obtain that:
clustering is scale and translation invariant and clusters do not tend to be of fixed size. Consequently we do not obtain the mouse effect\footnote{Let us add that in \cite{fahim2008k} the authors present a numerical modification of k-means to allow dealing with spherical shaped clusters of various size.}

%{\bf [Let us observe that in \cite{fahim2008k} one can find
%a modyfication of the k-means method which is able to discover 
%small clusters -- wiecej komentarza]}

%
\begin{example}
Let us observe on Figure \ref{fig:mouse} the comparison of Spherical CEC 
with classical k-means on the Mickey-Mouse-like set.
We see that Spherical CEC was able to find the ``right'' number of 
clusters, and that the clusters have ``reasonable'' shapes.
\end{example}

To apply Spherical clustering we need the same
information as in the classical k-means: in the case of k-means we seek the
splitting of the data $U \subset \R^N$ into $k$ sets $(U_i)_{i=1}^k$ such that the value of 
$
\sum \limits_{i=1}^k p(U_i) \cdot D_{U_i}
$
is minimal, where 
$D_{V}=\frac{1}{\card(V)}\sum \limits_{v \in V}\|v-\m_V\|^2$
denotes the mean within cluster $V$ sum of squares (and $\m_V$
is the mean of $V$). 
It occurs that the Gaussian spherical
clustering in $\R^N$ reduces to minimization of 
$$
\sum \limits_{i=1}^k p(U_i) \cdot [-\ln(p(U_i))+\frac{N}{2} \ln D_{U_i}].
$$

Next we proceed to the study of clustering by Gaussians with
fixed covariance. We show that in the case of bounded data
the optimal amount of clusters is bounded above by the
maximal cardinality of respective $\e$-net in the convex hull of the data.
We finish our paper with the study of clustering by Gaussian densities with covariance equal to $s \I$ and prove that with $s$ converging to zero we obtain the classical k-means clustering, while with $s$ growing to $\infty$ data will form one big group.

%% file: section/cross.tex
\section{Cross-entropy}

%%%%%%%%%%%%%%%%%%%%%%%%%%%%%%%%%%%%
\subsection{Compression and cross-entropy}

Since CEC is based on choosing the optimal
(from the memory point of view) coding algorithms, we first 
establish notation and present the basics of cross-entropy compression.

Assume that we are given a discrete probability distribution $\nu$ 
on a finite set $X=\{x_1,\ldots,x_k\}$ which attains the values $x_i$ with probabilities $f_i$. Then roughly speaking \cite{Co-Th} the optimal code-lengths\footnote{We accept arbitrary, not only integer, code-lengths.} in the case we use coding alphabet consisting of $d$ symbols to code $\nu$ are given by $l_i=-\log_d f_i$, and consequently the expected code length is given by the entropy
$$
\begin{array}{l}
h_d(\nu):=\sum \limits_{i=1}^k f_i l_i 
=\sum \limits_{i=1}^k f_i \cdot (-\log_d f_i)=
\frac{1}{\ln d}\sum \limits_{i=1}^k\sh(f_i), 
\end{array}
$$
where $\sh(x)$ denotes the Shannon function defined by $-x \cdot \ln x$ if $x>0$ and $\sh(0):=0$.
%Thus a probability distribution induces an, optimal from its ``memory point of view'', coding with lengths $l_i=-\log_d f_i$. 
We recall that for arbitrary code-lengths $l_i$ to be acceptable they have to satisfy Kraft's inequality
$
\sum \limits_{i=1}^k d^{-l_i} \leq 1
$.
%Given a sequence $(l_i)_{i=1}^k$ satisfying the Kraft's inequality, we can associate with it a subprobabilistic measure\footnote{The measure is subprobabilistic if the measure of the whole space is not greater then one.} $\nu:=\sum\limits_{i=1}^kf_i\d_{x_i}$, where $f_i=d^{-l_i}$. Let us observe that  as the length codes have to be integer, most codes in practice induce subprobabilistic measures.

If we code a discrete probability measure $\mu$ (which attains $x_i$ with probability $g_i$) by the code optimized for a subprobabilistic measure $\nu$ we arrive at the definition of {\em cross-entropy $h_d^{\times}(\mu\|f)$}, which is given as the expected code length
$$
h_d^{\times}(\mu\|f):=\sum_{i=1}^k g_i \cdot (-\log_d f_i).
$$

Let us proceed to the case of continuous probability measure $\nu$ on $\R^N$ (with density $f_\nu$). The role of entropy is played by {\em differential entropy} (which corresponds to the limiting value of discrete entropy of coding with quantization error going to zero \cite{Co-Th}):
$$
h_d(\nu):=\int f_\nu(x) \cdot (-\log_d f_\nu(x)) dx
=\frac{1}{\ln d}\int \sh(f_\nu(x))  dx,
$$
where $f_\nu$ denotes the density of the measure $\nu$. 
In fact, as was the case of discrete spaces, we will need to consider codings produced by subprobability measures.

\begin{definition}
We call a nonnegative measurable function $f:\R^N \to \R_+$ a {\em subdensity} if
$
\int_{\R^N} f(x)dx \leq 1
$.   
\end{definition}

Thus the differential code-length connected with subdensity $f$ is given by \begin{equation} \label{dile}
l(x)=-\log_d f(x). 
\end{equation}
Dually, an arbitrary measurable function $x \to l(x)$ is acceptable as a ``differential coding length'' if it satisfies the {\em differential version of the Kraft's inequality}:
\begin{equation} \label{diff-kraft}
\int d^{-l(x)} dx \leq 1,
\end{equation}
which is equivalent to saying that the function $f(x):=d^{-l(x)}$ is a subdensity.

From now on, if not otherwise specified, by $\mu$ we denote either continuous or discrete 
probability measure on $\R^N$.

\begin{definition}
We define the {\em cross-entropy of $\mu$ with respect to subdensity $f$}
by:
\begin{equation} \label{e0}
H_d^{\times}(\mu\|f):=\int -\log_d f(y) d\mu(y).
\end{equation}
\end{definition}

It is well-known that if $\mu$ has density $f_\mu$, the minimum in the above integral over all subdensities is obtained for $f=f_\mu$ (and consequently the cross-entropy is bounded from below by the differential entropy).

One can easily get the following:

\begin{observation} \label{reskala}
Let $f$ be a given subdensity and $A$ an invertible affine operation.
Then
$$
H^{\times}_d(\mu \circ A^{-1}\|f_A)=H^{\times}_d(\mu\|f)+
\log_d|\det A|,
$$
where $f_A$ is a subdensity defined by
\begin{equation} \label{fa}
f_A:x \to f(A^{-1}x)/|\det A|,
\end{equation}
and $\det A$ denotes the determinant of the linear component of $A$.
\end{observation}

In our investigations we will be interested in (optimal) coding for $\mu$ by elements of a set of subdensities $\F$, and therefore we put
$$
H_d^{\times}(\mu\|\F):=\inf \{H_d^{\times}(\mu\|f):f \in \F\}.
$$
One can easily check that if $\F$ consists of densities then the search for $H^{\times}(\mu\|\F)$ reduces to the maximum
likelihood estimation of measure $\mu$ by the family $\F$.
Thus by $\MLE(\mu\|\F)$ we will denote the set of all subdensities from $\F$ which realize the infimum:
$$
\MLE(\mu\|\F):=\{f \in \F: H_d^{\times}(\mu\|f)=H_d^{\times}(\mu\|\F)\}.
$$
In proving that the clustering is invariant with respect
to the affine transformation $A$ we will use the following simple corollary of Observation \ref{reskala}:

\begin{corollary} \label{obs:aff}
 Let $\F$ be the subdensity family and $A:\R^N \to \R^N$ 
an invertible affine operation. By $\F_A$ we denote $\{f_A: f \in \F\}$,
where $f_A$ is defined by \eqref{fa}.
Then 
\begin{equation}\label{obs:aff}
H_d^{\times}(\mu \circ A^{-1}\|\F_A)
=H_d^{\times}(\mu\|\F)+\log_d |\det A| .
\end{equation}
\end{corollary}

As we know entropy in the case when we code with $d$ symbols is a
rescaling of entropy computed in Nats, which can be symbolically written as:
$H_d(\cdot)=\frac{1}{\ln d}H_e(\cdot)$.
Consequently, for the shortness of notation when we 
compute the entropy in Nats we will omit the subscript $e$ in the entropy symbol 
$$
H^{\times}(\cdot):=H_e^{\times}(\cdot).
$$

%%%%%%%%%%%%%%%%%%%%%%%%%%%%%%%%%%%
\subsection{Cross-entropy for Gaussian families}

By $\m_{\mu}$ and $\Sigma_{\mu}$ we denote the mean and covariance of the measure $\mu$, that is
$$
\begin{array}{l}
\m_{\mu}=\frac{1}{\mu(\R^N)}\int x \, d\mu(x), \\[1ex]
\Sigma_{\mu}=\frac{1}{\mu(\R^N)}\int (x-\m_{\mu})(x-\m_{\mu})^T \, d \mu(x).
\end{array}
$$
For measure $\mu$ and measurable set $U$ such that $0<\mu(U)<\infty$ we introduce the probability measure $\mu_U$
$$
\mu_U(A):=\mbox{$\frac{1}{\mu(U)}$}\mu(A \cap U),
$$
and use the abbreviations
$$
\begin{array}{l}
\m_U^{\mu}:=\m_{\mu_U}=\frac{1}{\mu(U)}\int_{U} x \, d\mu(x), \\[1ex]
\Sigma_U^{\mu}:=\Sigma_{\mu_U}=\frac{1}{\mu(U)}\int_{U} (x-\m_{\mu})(x-\m_{\mu})^T \, d \mu(x).
\end{array}
$$

The basic role in Gaussian cross-entropy minimization is played by the 
following result which says that we can reduce computation to gaussian families. Since its proof is essentially known part of MLE, we provide here only its short idea.

Given symmetric positive matrix $\Sigma$, we recall that
by the Mahalanobis distance \cite{Ma, Ma2} we understand
$$
\|x-y\|_{\Sigma}:=(x-y)^T\Sigma^{-1}(x-y).
$$
By $\Nor{(\m,\Sigma)}$ we denote the normal density with
mean $\m$ and covariance $\Sigma$, which as we recall is given by
the formula
$$
\Nor{(\m,\Sigma)}(x):=\frac{1}{\sqrt{(2\pi)^N\det(\Sigma)}} \exp(\mbox{$\frac{1}{2}$}\|x-\mu\|_\Sigma^2).
$$

\begin{theorem} \label{pr1}
Let $\mu$ be a discrete or continuous probability measure with well-defined covariance matrix,
and let $\m \in \R^N$ and positive-definite symmetric matrix $\Sigma$ be given.

Then
$$
\ce{\Nor{(\m,\Sigma)}}=H^{\times}\big(\mu_{\G}\|\Nor{(\m,\Sigma)}\big),
$$
where $\mu_\G$ denotes the probability measure with Gaussian density
of the same mean and covariance as $\mu$ (that is the density of $\mu_\G$ equals $\Nor(\m_{\mu},\Sigma_{\mu})$).

Consequently
\begin{equation} \label{e2}
%\begin{array}{l}
\ce{\Nor{(\m,\Sigma)}}=\frac{N}{2} \ln(2\pi)+
% \\[1ex]\phantom{=}
\frac{1}{2}\|\m-\m_{\mu}\|^2_{\Sigma}+\frac{1}{2}\tr(\Sigma^{-1}\Sigma_{\mu})+\frac{1}{2}\ln \det(\Sigma).
%\end{array}
\end{equation}
\end{theorem}

\begin{proof}[Sketch of the proof]
We consider the case when $\mu$ is a continuous measure with density $f_\mu$.
One can easily see that by applying trivial affine transformations and
\eqref{obs:aff} it is sufficient to prove \eqref{e2} in the case when $\m=0$ and $\Sigma=\I$. Then we have
$$
\begin{array}{l}
H^{\times}(\mu\|\Nor{(0,\I)}) %\\[1ex]
=  \int f_\mu(x) \cdot [\frac{N}{2} \ln(2\pi)+\frac{1}{2}\ln \det(\I)+\frac{1}{2}\|x\|^2]dx \\[1ex]
=  \frac{N}{2} \ln(2\pi)+\frac{1}{2} \int f_\mu(x) \|(x-\m_{\mu})+\m_{\mu}\|^2 dx \\[1ex]
 =  \frac{N}{2} \ln(2\pi) + \!\frac{1}{2} \int \!f_\mu(x) [\|x-\m_{\mu}\|^2\! +\|\m_{\mu}\|^2\!+2(x-\m_{\mu}) \circ \m_{\mu}] dx \\[1ex] 
 =  \frac{N}{2} \ln(2\pi)+\frac{1}{2}\tr(\Sigma_{\mu})+\frac{1}{2}\|\m_{\mu}\|^2.
\end{array}
$$
\end{proof}

By $\G$ we denote the set of all normal densities, while by $\G_\Sigma$ we denote the set of all normal densities with covariance $\Sigma$.
As a trivial consequence of the Theorem \ref{pr1} we obtain the following
proposition.

\begin{proposition} \label{fis}
Let $\Sigma$ be a fixed positive symmetric matrix. Then $\MLE(\mu\|\G_{\Sigma})=\{\Nor{(\m_{\mu},\Sigma)}\}$ and
$
H^{\times}(\mu\|\G_{\Sigma})=
\frac{N}{2}\ln(2\pi)+\frac{1}{2}\tr(\Sigma^{-1}\Sigma_{\mu})+\frac{1}{2}
\ln \det(\Sigma)
$.
\end{proposition}

Now we consider cross-entropy with respect to all normal
densities.

\begin{proposition} \label{cross-gaussians}
We have $\MLE(\mu\|\G)=\{\Nor{(\m_{\mu},\Sigma_\mu)}\}$ and
\begin{equation}
H^{\times}(\mu\|\G)=\frac{1}{2}\ln \det(\Sigma_{\mu})+\frac{N}{2}\ln(2\pi e).
\end{equation}
\end{proposition}

\begin{proof}
Since entropy is minimal when we code a measure by its own density, we easily
obtain that 
$$
H^{\times}(\mu\|\G)=H^{\times}(\mu_\G\|\G)
=H^{\times}(\mu_\G\|\Nor(\m_\mu,\Sigma_\mu))
$$
$$
=H(\mu_\G)
=\frac{1}{2}\ln \det(\Sigma_{\mu})+\frac{N}{2}\ln(2\pi e).
$$
Consequently the minimum is realized for $\Nor{(\m_\mu,\Sigma_{\mu})}$. 
\end{proof}

Due to their importance and simplicity we also consider Spherical Gaussians $\S$, that is those with covariance matrix proportional to $\I$:
$$
\S=\bigcup_{s>0} \G_{s\I}.
$$
We will need the denotation for the mean squared distance from the mean\footnote{We will see that it corresponds to the mean within clusters some of squares.}
$$
D_{\mu}:=\int \|x-\m_{\mu}\|^2d\mu(x)=\tr (\Sigma_{\mu}),
$$
which will play in Spherical Gaussians the analogue of covariance.
As is the case for the covariance, we will use the abbreviation
$$
D_U^\mu:=D_{\mu_U}=\frac{1}{\mu(U)}
\int_U \|x-\m^\mu_U\|^2d\mu(x).
$$

Observe, that if $\Sigma_{\mu}=s \I$ then $D_{\mu}=Ns$.
In the case of one dimensional measures $\sqrt{D_{\mu}}$ is
exactly the standard deviation of the measure $\mu$.
It occurs that $D_{\mu}$
can be naturally interpreted as a square of the ``mean radius'' of the measure $\mu$:  for the uniform probability measure $\mu$ on the sphere $S(x,R) \subset \R^N$ we clearly get $\sqrt{D_{\mu}}=R$. 
Moreover, as we show in the following observation  $\sqrt{D_{\mu}}$ will be close to $R$ even for the uniform probability distribution on the ball $B(x,R)$.

By $\lambda$ we denote the Lebesgue measure on $\R^N$. Recall that according to our notation $\lambda_U$ denotes the probability measure
defined by $\lambda_U(A):=\lambda(A \cap U)/\lambda(U)$.

\begin{observation}
We put $V_n:=\lambda(B_n(0,1))=\pi^{n/2}/\Gamma(n/2+1)$, where $B_n(0,1)$
denotes the unit ball in $\R^n$.

Consider the unit ball $B(0,1) \subset \R^N$. Directly from the definition of covariance we get $\Sigma_{B(0,1)}^{\lambda}=c_N \I$, where 
$$
\begin{array}{l}
c_N=\frac{1}{V_N}\int \limits_{-1}^1 x^2 V_{N-1}
\cdot (\sqrt{1-x^2})^{N-1} dx \\
=\frac{\Gamma(N/2+1)}{\sqrt{\pi}\Gamma((N-1)/2+1)}\int \limits_{-1}^1
x^2 (1-x^2)^{(N-1)/2}dx=\frac{1}{N+2}.
\end{array}
$$
Consequently, 
\begin{equation} \label{wazne}
\begin{array}{l}
\Sigma^{\lambda}_{B(x,R)}=R^2 \Sigma^{\lambda}_{B(0,1)}=\frac{R^2}{N+2}, \\[1ex]
D^{\lambda}_{B(x,R)}=
\tr (\Sigma^\lambda_{B(x,R)})=\frac{NR^2}{N+2},
\end{array}
\end{equation}
and therefore $
\sqrt{D^\lambda_{B(x,R)}}=\sqrt{\frac{N}{N+2}} \cdot R \to R \mbox{ as }N \to \infty$.
\end{observation}

\begin{proposition} \label{cross-spherical}
We have $\MLE(\mu\|\S)=\{\Nor(\m_{\mu},\frac{D_\mu}{N}\I)\}$
and
\begin{equation} \label{cd}
H^{\times}(\mu\|\S)=\frac{N}{2}\ln (D_\mu)+\frac{N}{2}\ln(2\pi e/N).
\end{equation}
\end{proposition}

\begin{proof}
Clearly by Proposition \ref{fis}
$$
H^{\times}(\mu\|\S)=\inf_{s>0}
H^{\times}(\mu\|\G_{s\I})
=\inf_{s>0} \left(\frac{1}{2s}D_{\mu}+\frac{N}{2}\ln s+\frac{N}{2}\ln(2\pi)\right).
$$
Now by easy calculations we obtain that the above function attains
minimum for $s=D_{\mu}/N$ and equals the RHS of \eqref{cd}.
\end{proof}

%\begin{example}
%Let $R[r]$ denote the rectangle $[0,r] \times [0,1/r]$. Then the 
%covariance matrix of $\lambda_{R[r]}$ is given by
%$$
%\Sigma^\lambda_{R[r]}=
%\frac{1}{12}\left[
%\begin{array}{cc}
%r^2 &0 \\
%0 & 1/r^2
%\end{array}
%\right].
%$$
%and consequently by Propositions \ref{cross-gaussians} and \ref{cross-spherical}
%$$
%\begin{array}{l}
%H^{\times}(\lambda_{R[r]}\|\G)=\ln(\frac{\pi e}{6}),\\
%H^{\times}(\lambda_{R[r]}\|\S)=\ln(\frac{\pi e}{6})+\ln(\frac{r^2+1/r^2}{2}).
%\end{array}
%$$
%Thus as expected, the coding of rectangles by arbitrary gaussian densities is better then coding by spherical gaussians, with the difference
%going to $\infty$ as $r$ goes to zero or infinity. 
%\end{example}

At the end we consider the cross-entropy with respect to
$\G_{s\I}$ (spherical Gaussians with fixed scale). As
a direct consequence of Proposition \ref{fis} we get:

\begin{proposition}
Let $s>0$ be given. Then $\MLE(\mu\|\G_{\s\I})=\{\Nor(\m_\mu,s\I)\}$
and
$$
H^{\times}(\mu\|\G_{\s\I})
=\frac{1}{2s}D_{\mu}+\frac{N}{2}\ln s+\frac{N}{2}\ln(2\pi).
$$
\end{proposition}

%\begin{example}
%Let us  consider the cross-entropy of the uniform probability
%measure on the unit ball $B(0,1) \subset \R^N$. By \eqref{wazne}
%we know that $D^\lambda_{B(0,1)}=\frac{N}{N+2}$. 
%which implies that
%$$
%\begin{array}{l}
%H^{\times}(\lambda_{B(0,1)}\|\G_{(\cdot \I)})
%=\frac{N}{2}(1- \ln(N+2))+\frac{N}{2}\ln(2\pi), \\[1ex]
%H^{\times}(\lambda_{B(0,1)}\|\G_{s\I})=
%\frac{N}{2}(\frac{1}{(N+2)s}+\ln s)+\frac{N}{2}\ln(2\pi).
%\end{array}
%$$
%Observe that if $s$ is sufficiently far from the optimal value $1/(N+2)$, i.e.
%either sufficiently small or sufficiently large,
%the difference between the cross-entropies can be arbitrarily large.
%\end{example}

%% file: section/many.tex
\section{Many coding subdensities}

%%%%%%%%%%%%%%%%%%%%%%%%%%%%%%%%%%%%
\subsection{Basic idea}

In the previous section we considered the coding with $d$-symbols of the
$\mu$-randomly chosen point $x \in \R^N$ by the code optimized for the subdensity $f$. 
Since it is often better to ``pack/compress''  
parts of data with various algorithms, we follow this approach 
and assume that we are given a sequence of $k$ 
subdensities\footnote{In general we accept also $k=\infty$.} $(f_i)_{i=1}^k$, which we interpret as coding algorithms. 

Suppose that we want to code $x$ by $j$-th algorithm from the sequence $(f_i)_{i=1}^k$. By \eqref{dile} the length of code of $x$ corresponds to
$-\log_d f_j(x)$. However, this code itself is clearly insufficient to
decode $x$ if we do not know which coding algorithm was used.
Therefore to uniquely code $x$ we have to add to it the code of $j$. Thus if $l_j$ denotes the length of code of $j$, the ``final'' length of the code of the point $x$ is the sum of $l_j$ and the length of the code of the point $x$:
$$
\mbox{code-length of }x=l_j-\log_d f_j(x).
$$
Since the coding of the algorithms has to be acceptable, the sequence $(l_i)_{i=1}^k$ has to satisfy the Kraft's inequality and therefore if we put $p_i=d^{-l_i}$, we can consider only those $p_i \geq 0$ that $\sum \limits_{i=1}^k p_i \leq 1$. Consequently without loss of generality (by possibly shortening the expected code-length), we may restrict to the case when $\sum \limits_{i=1}^kp_i=1$.

Now suppose that points from $U_i \subset \R^N$ we code by the subdensity $f_i$. Observe that although $U_i$ have to be pairwise disjoint, they
do not have to cover the whole space $\R^N$ -- we can clearly
omit the set with $\mu$-measure zero. To formalize this, the notion of $\mu$-partition\footnote{We introduce $\mu$-partition as in dealing in practice with clustering of the discrete data
it is natural to partition just the dataset and not the whole space.} for 
a given continuous or discrete measure $\mu$ is convenient: we say that
a pairwise disjoint sequence $(U_i)_{i=1}^k$ of Lebesgue measurable subsets of $\R^N$ is a {\em $\mu$-partition} if 
$$
\mu \left( \R^N \setminus \bigcup_{i=1}^k U_i \right) =0.
$$

{\em To sum up:} we have the ``coding'' subdensities $(f_i)_{i=1}^k$ and $p \in P_k$, where
$$
P_k:=\{(p_1,\ldots,p_k) \in [0,1]^k :\sum_{i=1}^kp_i=1\}.
$$
As $U_i$ we take the set of points of $\R^N$ we code by density $f_i$. 
Then for a $\mu$-partition $(U_i)_{i=1}^k$ we obtain the code-length function 
$$
x \to -\log_d p_i-\log_d f_i(x) \for x \in U_i,
$$ 
which is exactly the code-length of the subdensity\footnote{Observe
that this density is defined for $\mu$-almost all $x \in \R^N$.}
\begin{equation} \label{e-dens}
p_1f_1|_{U_1} \cup \ldots \cup p_kf_k|_{U_k}.
\end{equation}
In general we search for those $p$ and $\mu$-partition for which
the expected code-length given by the cross-entropy 
$
\ce{\ocup \limits_{i=1}^k p_if_i|_{U_i}}
$
will be minimal.

\begin{definition}
Let $(\F_i)_{i=1}^k$ be a sequence of subdensity families in $\R^N$, and
let a $\mu$-partition $(U_i)_{i=1}^k$ be given.
Then we define
$$
\oti \limits_{i=1}^k (\F_i|U_i):=
\big\{\ocup \limits_{i=1}^k p_if_i|_{U_i} :
(p_i)_{i=1}^k \in P_k, (f_i)_{i=1}^k \in (\F_i)_{i=1}^k \big\}.
$$
\end{definition}

Observe that 
$
\oti \limits_{i=1}^k (\F_i|U_i)
$
denotes those compression algorithms which can be build by using
an arbitrary compression subdensity from $\F_i$ on the set $U_i$.

%%%%%%%%%%%%%%%%%%%%%%%%%%

\subsection{Lloyd's algorithm}

The basic aim of our article is to 
find a $\mu$-partition $(U_i)_{i=1}^k$
for which 
$$
\ced{\oti \limits_{i=1}^k (\F_i|U_i)}
$$
is minimal. In general it is NP-hard problem even for k-means \cite{NPhard}, which is the simplest limiting case of Spherical CEC (see Observation \ref{obs-kmeans}). However, in practice we can often hope to find
a sufficiently good solution by applying either Lloyd's or Hartigan's method.

The basis of Lloyd's approach is given by the following
two results which show that 
\begin{itemize}
\item given $p \in P_k$ and $(f_i)_{i=1}^k \in (\F_i)_{i=1}^k$, we can find a partition $(U_i)_{i=1}^k$ which minimizes the cross-entropy $\ce{\ocup \limits_{i=1}^k p_if_i|_{U_i}}$;
\item for a partition $(U_i)_{i=1}^k$, we can find $p \in P_k$ and $(f_i)_{i=1}^k \in (\F_i)_{i=1}^k$  which minimizes $\ce{\ocup \limits_{i=1}^k p_if_i|_{U_i}}$.
\end{itemize}

We first show how to minimize the value of cross-entropy 
being given a $\mu$-partition $(U_i)_{i=1}^k$.
From now on we interpret $0 \cdot x$ as zero even if
$x=\pm \infty$ or $x$ is not properly defined.

\begin{observation} \label{great}
Let
$(f_i) \in (\F_i)$, $p \in P_k$ and $(U_i)_{i=1}^k$ be a $\mu$-partition. Then
\begin{equation} \label{formu}
\ce{\ocup \limits_{i=1}^k p_if_i|_{U_i}} 
=\sum \limits_{i=1}^k \mu(U_i)\cdot \left(-\ln p_i+H^{\times}(\mu_{U_i}\|f_i) \right).
\end{equation}
\end{observation}

\begin{proof}
We have
$$
\begin{array}{rcl}
\ce{\ocup \limits_{i=1}^k p_if_i|_{U_i}} 
& = &\sum \limits_{i=1}^k \int_{U_i} -\ln p_i-\log_d f_i(x) d\mu(x) \\[0.5ex]
& = &\sum \limits_{i=1}^k \mu(U_i)\cdot \left(-\ln p_i-\int \ln (f_i(x)) d\mu_{U_i}(x)
\right).
\end{array}
$$
\end{proof}

\begin{proposition} \label{pop}
Let the sequence of subdensity families $(\F_i)_{i=1}^k$ be given
and let $(U_i)_{i=1}^k$ be a fixed $\mu$-partition. 
We put $p=(\mu(U_i))_{i=1}^k \in P_k$.

Then
$$
\ce{\oti_{i=1}^k (\F_i|U_i)}=\ce{\ocup \limits_{i=1}^k p_if_i|_{U_i}}
$$
$$
=\sum_{i=1}^k \mu(U_i) \cdot [-\ln(\mu(U_i)+H^{\times}(\mu_{U_i}\|\F_i)].
$$
\end{proposition}

\begin{proof}
We apply the formula \eqref{formu}
$$
\ce{\ocup \limits_{i=1}^k \tilde p_if_i|_{U_i}}
=\sum \limits_{i=1}^k \mu(U_i)\cdot \left(-\ln \tilde p_i+H^{\times}(\mu_{U_i}\|f_i) \right).
$$
By the property of classical entropy we know that the function
$$
P_k \ni \tilde p=(\tilde p_i)_{i=1}^k \to \sum \limits_{i=1}^k \mu(U_i)\cdot (-\ln \tilde p_i)
$$ 
is minimized for $\tilde p=(\mu(U_i))_i$.
\end{proof}

The above can be equivalently rewritten with the use of notation:
$$
h_{\mu}(\F;W):=
\begin{cases}\mu(W) \!\cdot \!\big(-\ln(\mu(W))+H^{\times}(\mu_W\|\F)\big)
\mbox{ if } \mu(W)>0,\\
0 \mbox{ otherwise.}
\end{cases}
$$
Thus $h_\mu(\F;W)$ tells us what is the minimal cost of compression of the part of our dataset contained in $W$ by subdensities
from $\F$. By Proposition \ref{pop} if $(U_i)_{i=1}^k$ is 
a $\mu$-partition then
\begin{equation} \label{krent}
\ce{\oti_{i=1}^k (\F_i|U_i)}=\sum_{i=1}^k h_{\mu}(\F_i;U_i).
\end{equation}
Observe that, in general, if $\mu(U)>0$ then
$$
H^{\times}(\mu_U\|\F)=\ln(\mu(U))+\frac{1}{\mu(U)}h_\mu(\F;U).
$$
Consequently, if we are given a $\mu_U$-partition $(U_i)_{i=1}^k$, then
$$
H^{\times}(\mu_U\|\oti \limits_{i=1}^k(\F_i|U_i))=
\ln(\mu(U))+\frac{1}{\mu(U)}\sum \limits_{i=1}^k h_\mu(\F_i;U_i).
$$

\begin{theorem} \label{llo}
Let the 
sequence of subdensity families $(\F_i)_{i=1}^k$ be given
and let $(U_i)_{i=1}^k$ be a fixed $\mu$-partition. 

We put $p=(\mu(U_i))_{i=1}^k \in P_k$.
We assume that $\MLE(\mu_{U_i}\|\F_i)$ is nonempty for every $i=1..k$.
Then for arbitrary
\begin{equation} \label{dd}
f_i \in \MLE(\mu_{U_i}\|\F_i) \for i=1,\ldots,k,
\end{equation}
we get
$$
\ce{\oti_{i=1}^k(\F_i|U_i)}=
\ce{\oti_{i=1}^k p_i f_i|_{U_i}}.
$$
\end{theorem}

\begin{proof}
Directly from the definition of $\MLE$ we obtain that
$$
H^{\times}(\mu_{U_i}\|\tilde f_i)  \geq  H^{\times}(\mu_{U_i}\|\F_i)
=H^{\times}(\mu_{U_i}\|f_i)
$$
for $\tilde f_i \in \F_i$.
\end{proof}

The following theorem is a dual version of Theorem \ref{llo} -- 
for fixed $p \in P_k$ and $f_i \in \F_i$ we seek optimal
$\mu$-partition which minimizes the cross-entropy.

By the support of measure $\mu$ we denote the support of its
density if $\mu$ is continuous and the set of support points if 
it is discrete.

\begin{theorem} \label{llo2}
Let the 
sequence of subdensity families $(\F_i)_{i=1}^k$ be given
and let $f_i \in \F_i$ and $p \in P_k$ be such that
$\supp(\mu) \subset \bigcup \limits_{i=1}^k \supp (f_i)$. 
We define $l:\supp(\mu) \to (-\infty,\infty]$ by
$$
l(x):=\min \limits_{i \in \{1,\ldots,k\}}[-\ln p_i-\ln f_i(x)].
$$
 
We construct a sequence $(U_i)_{i=1}^k$ of measurable subsets of $\R^N$ recursively by the following procedure:
\begin{itemize}
\item $U_1=\{x \in \supp(\mu)  : -\ln p_1-\ln f_1(x)=l(x)
\}$;
\item $U_{l+1}=\{x \in \supp(\mu) \setminus (U_1 \cup \ldots \cup U_l) 
 : -\ln p_{l+1}-\ln f_{l+1}(x)=l(x)\}$.
\end{itemize}
Then $(U_i)_{i=1}^k$ is a $\mu$-partition and
$$
\ce{\ocup \limits_{i=1}^k p_if_i|_{U_i}}=\inf\{\ce{\ocup \limits_{i=1}^k p_if_i|_{V_i}}
: \mbox{$\mu$-partition }(V_i)_{i=1}^k \}.
$$
\end{theorem}

\begin{proof}
Since $\supp(\mu) \subset \bigcup \limits_{i=1}^k \supp(f_i)$, we
obtain that $(U_i)_{i=1}^k$ is a $\mu$-partition.
Moreover, directly by the choice of $(U_i)_{i=1}^k$ we obtain that
$$
\begin{array}{l}
l(x)=\ln(\ocup \limits_{i=1}^k p_if_i|_{U_i})(x) \for x \in \supp(\mu),
\end{array}
$$
and consequently 
for an arbitrary $\mu$-partition $(V_i)_{i=1}^k$ we get
$$
\begin{array}{l}
H^{\times}(\mu\|\ocup \limits_{i=1}^k p_if_i|_{V_i}) 
 = \int \ocup \limits_{i=1}^k \big[-\ln (p_i)-\ln(f_i|_{V_i}(x)) \big]  d\mu(x) \\[0.5ex]
 \leq  \int \bar l(x)  d\mu(x) 
=\int \ocup \limits_{i=1}^k \big[-\ln (p_i)-\ln(f_i|_{U_i}(x)) \big]  d\mu(x).
\end{array}
$$
\end{proof}

As we have mentioned before, Lloyd's approach is based on alternate use of steps from Theorems \ref{llo} and \ref{llo2}. In practice we usually start by choosing initial
densities and set probabilities $p_k$ equal: $p=(1/k,\ldots,1/k)$ 
(since the convergence is to local minimum we commonly start from various initial condition several times).

%\medskip
%
%\begin{algorithm}                      % enter the algorithm environment
%\caption{(\bf LLOYD'S ALGORITHM):}          % give the algorithm a caption
%\label{alg0}                           % and a label for \ref{} commands later in the document
%\begin{algorithmic}                    % enter the algorithmic environment
%\footnotesize
%%
%\STATE {\bf put}
%\STATE $n:=0$, $h_0:=\infty$, $p:=(1/k,\ldots,1/k)$, $U_0=\emptyset$
%\STATE {\bf choose}
%\STATE $\e>0$
%\STATE $(f_i)_{i=1}^k \in (\F_i)_{i=1}^k:\supp(\mu) \subset \bigcup_{i=1}^k \supp(f_i)$
%\REPEAT
%\STATE $n=n+1$
%\STATE $l(x)=\min \limits_{i \in \{1,\ldots,k\}}[-\ln p_i-\ln f_i(x)]$
%\FOR{ $i=1:k$ }
%\STATE $U_i=\{x \in \supp(\mu) \setminus (U_0 \cup \ldots \cup U_{i-1}) 
% : -\ln p_i-\ln f_i(x)=l(x)\}$
%\ENDFOR
%\STATE $p=(\mu(U_i))_{i=1}^k$
%\FOR{$i=1:k$}
%\STATE {\bf compute} $f_i \in \MLE(\mu_{U_i}\|\F_i)$
%\ENDFOR
%\STATE $h_n:=\sum \limits_{i=1}^k[\sh(\mu(U_i))-\int_{U_i}\ln f_i d\mu]$
%\UNTIL{ $h_n \geq h_{n-1}-\e$ }
%\end{algorithmic}
%\end{algorithm}
%
%\bigskip

Observe that directly by Theorems \ref{llo} and \ref{llo2} we obtain that the
sequence $n \to h_n$ is decreasing. One 
hopes\footnote{To enhance that chance we usually start many
times from various initial clustering.} that limit $h_n$ converges (or at least is reasonably close) to the global infimum of $\ce{\oti \limits_{i=1}^k(\F_i|U_i)}$. 

\medskip

To show a simple example of cross-entropy minimization
we first need some notation. 
We are going to discuss the Lloyds cross-entropy minimization of discrete data 
with respect to $\G_{\Sigma_1},\ldots,\G_{\Sigma_K}$.
As a direct consequence of \eqref{krent} and Proposition \ref{fis} we obtain
the formula for the cross entropy of $\mu$ with respect to a family
of Gaussians with covariances $(\Sigma_i)_{i=1}^k$.

\begin{observation}
Let $(\Sigma_i)_{i=1}^k$ be fixed positive symmetric matrices
and let $(U_i)_{i=1}^k$ be a given $\mu$-partition.
Then
$$
\begin{array}{l}
H^{\times}(\mu\|\oti\limits_{i=1}^k(\G_{\Sigma_i}|U_i)) =
\\[1ex]
\frac{N}{2} \ln(2\pi)+\sum \limits_{i=1}^k \mu(U_i) \big[\!-\ln(\mu(U_i))+\frac{1}{2}\tr(\Sigma_i^{-1}\Sigma^{\mu}_{U_i})+\frac{1}{2}
\ln \det(\Sigma_i)\big].
\end{array}
$$
\end{observation}
 
\begin{example}
\begin{figure}[!t]
\centering
	\subfigure[$T$-like set.]{\label{fig:v_1_a}
\fbox{\includegraphics[width=1.4in]{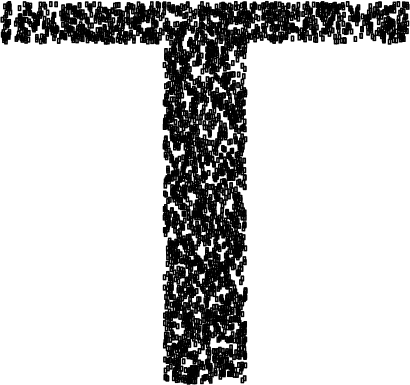}}}
	\subfigure[Partition into two groups.]{\label{fig:v_2_a}
\fbox{\includegraphics[width=1.4in]{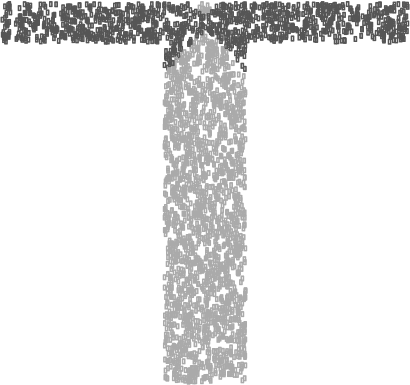}}}
	\caption{Effect of clustering by Lloyd's algorithm on $T$-like shape.}
	\label{fig:t} 
\end{figure}
We show Lloyd's approach to cross-entropy
minimization of the set $\y$ showed on Figure \ref{fig:v_1_a}.
As is usual, we first associate with the data-set $\y$ the probability measure
defined by the formula
$$
\mu:=\frac{1}{\card \y} \sum_{y \in \y} \d_y,
$$
where $\d_y$ denotes the Dirac delta at the point $y$.

Next we search for the $\mu$-partition $\y=\y_1 \sqcup \y_2$
which minimizes
$$
\ce{(\G_{\Sigma_1}|\y_1) \otis (\G_{\Sigma_2}|\y_2)},
$$
where $\Sigma_1=[300,0;0,1]$, $\Sigma_2=[1,0;0,300]$.
The result is given on Figure \ref{fig:v_2_a}, where the dark gray points
which belong to $\y_1$ are ``coded'' by density from
$\G_{\Sigma_1}$ and light gray belonging to $\y_2$ and are ``coded''
by density from $\G_{\Sigma_2}$.
\end{example}

%%%%%%%%%%%%%%%%%%%%%%%%%%%%%%%%%%%
\subsection{Hartigan algorithm}

Due to its nature to use Hartigan we have to divide the data-set 
(or more precisely the support of the measure $\mu$) into ``basic  parts/blocks'' from which we construct our clustering/grouping.  Suppose that we have a fixed $\mu$-partition\footnote{By default we think of it as a partition into sets with small diameter.} $\V=(V_i)_{i=1}^n$. The aim of Hartigan is to find
such $\mu$-partition build from elements of $\V$ which has minimal cross-entropy. 

Consider $k$ coding subdensity families $(\F_i)_{i=1}^k$.
To explain Hartigan approach more precisely we need the notion of {\em group 
membership function} $\cl:\{1,\ldots,n\} \to \{0,\ldots,k\}$ which 
describes the membership of $i$-th element of partition, where $0$
value is a special symbol which denotes that $V_i$ is as yet unassigned.
In other words: if $\cl(i)=l>0$, then $V_i$ is a part of the $l$-th group, and
if $\cl(i)=0$ then $V_i$ is unassigned.

We want to find such $\cl:\{1,\ldots,n\} \to \{1,\ldots,k\}$ (thus all elements of $\V$ are assigned) that
$$
\sum_{i=1}^k h_\mu(\F_i;\V(\cl^{-1}(i))
$$ 
is minimal. Basic idea of Hartigan is relatively simple -- we repeatedly go over all elements of the partition $\V=(V_i)_{i=1}^n$ and apply the following steps:
\begin{itemize}
\item if the chosen set $V_i$ is unassigned, assign it to the first nonempty group;
\item reassign $V_i$ to those group for which the decrease in cross-entropy is maximal;
\item check if no group needs to be removed/unassigned, if this is the case unassign its all elements;
\end{itemize}
until no group membership has been changed.

To practically apply Hartigans algorithm we still have to decide about the way we choose initial group membership.
In most examples in this paper we initialized the cluster membership function randomly. However, one can naturally speed the clustering by using some more intelligent
cluster initialization which are often commonly encountered in the 
modifications of k-means (one can for example easily use k-means++ approach \cite{kmeans++}).

To implement Hartigan approach for discrete measures we still have to add a condition when we unassign given group. For example in the case of Gaussian clustering in $\R^N$ to avoid overfitting we cannot consider clusters which contain less then $N+1$ points. In practice while applying Hartigan approach on discrete data we usually removed clusters which contained less then three percent of all data-set. 

Observe that in the crucial step in Hartigan approach we compare the cross entropy after and before the switch, while the switch removes a given set from one cluster and adds it to the other. Since 
$$
h_\mu(\F;W)=\mu(W)\cdot \big(-\ln(\mu(W))+H^{\times}(\mu_W\|\F) \big),
$$
basic steps in the Hartigan approach reduce to computation of $H^{\times}(\mu_W\|\F))$
for $W=U \cup V$ and $W=U \setminus V$.
This implies that to apply efficiently the Hartigan approach in clustering
it is of basic importance to compute 
\begin{itemize}
\item $H^{\times}(\mu_{U \cup V}\|\F)$ for disjoint $U,V$;
\item $H^{\times}(\mu_{U \setminus V}\|\F)$ for $V \subset U$.
\end{itemize}
%To illustrate it,
%let us first investigate the above on the simple case of family consisting
%from one density.
%
%\begin{example}
%Let $\F=\{f\}$, where $f$ is a fixed density and let $U \subset \supp(f)$.
%Then $H^{\times}(\mu_U\|\F)=\frac{1}{\mu(U)}\int_U -\ln(f(x))d\mu(x)$.
%Consequently for $V \subset \supp(f)$ disjoint with $U$ we have
%$$
%H^{\times}(\mu_{U \cup V}\|\F)=
%\frac{\mu(U)}{\mu(U)+\mu(V)}H^{\times}(\mu_U\|\F)+\frac{\mu(V)}{\mu(U)+\mu(V)}H^{\times}(\mu_V\|\F).
%$$ 
%On the other hand, for $V \subset U$ such that $\mu(V) <\mu(U)$
%we have
%$$
%H^{\times}(\mu_{U \setminus V}\|\F)=\\[0.5ex]
%\frac{\mu(U)}{\mu(U)-\mu(V)}H^{\times}(\mu_U\|\F)-\frac{\mu(V)}{\mu(U)-\mu(V)}H^{\times}(\mu_V\|\F).
%$$
%\end{example}
%
%Now we are going to deal with similar question for Gaussian families.
Since in the case of Gaussians to compute the cross-entropy of $\mu_W$ we need only covariance $\Sigma_W^{\mu}$, our problem reduces to computation of $\Sigma_{U \cup V}$ and $\Sigma_{U \setminus V}$.
%We are going to addresses
%this issue.
% We recall that for a probability measure we have
%$$
%\begin{array}{l}
%\m_{\mu}:=\int x \, d\mu(x), \\[1ex]
%\Sigma_{\mu}:=\int (x-\m_\mu)(x-\m_{\mu})d\mu(x)
%%\\[0.5ex] \phantom{\Sigma_{\mu}}
%=\int x \cdot x^T \, d\mu(x)-\m_{\mu} \cdot \m_{\mu}^T. 
%\end{array}
%$$
%Let us start with the following easy theorem
%which says what is the mean and covariance of convex combinations
%of measures.
%\begin{proposition}
%Let $p \in P_k$ and $(\mu_i)_{i=1}^k$ be a sequence probability measures on $\R^N$ for which the mean and covariance are well-defined.
%
%We put $\m_i=\m_{\mu_i}$, $\Sigma_i=\Sigma_{\mu_i}$ and $\mu=\sum \limits_{i=1}^k p_i\mu_i$. Then
%$$
%\begin{array}{l}
%\m_{\mu}  =  \sum \limits_{i=1}^k p_i \m_i, \mbox{ and }
%\Sigma_{\mu} 
% =  \sum \limits_{i=1}^k p_i \Sigma_i+\sum \limits_{i=1}^k p_i (\m_i-\m_{\mu})(\m_i-\m_{\mu})^T. 
%\end{array}
%$$
%\end{proposition}
%
%\begin{proof}
%We have
%$$
%\begin{array}{l}
%\Sigma_{\mu}=\int xx^Td\mu(x)-\m_\mu\m_\mu^T %\\[1ex]
%=\sum \limits_{i=1}^k p_i(\Sigma_i+\m_i\m_i^T)-\m_\mu\m_\mu^T.
%\end{array}
%$$
%\end{proof}
%
%Consequently, 
One can easily verify that for %the most important case of 
convex combination of two measures we have:
             
\begin{theorem} \label{lop}
Let $U,V$ be Lebesgue measurable sets with finite and nonzero $\mu$-measures. 

a) Assume additionally that $U \cap V=\emptyset$. Then 
$$
\begin{array}{l}
\m^{\mu}_{U \cup V} =  p_{U}\m^{\mu}_{U}+p_V\m^{\mu}_V, \\[1ex]
\Sigma^{\mu}_{U \cup V} =  
%\\\phantom{=}
p_U\Sigma^{\mu}_U+p_V\Sigma^{\mu}_V
+p_Up_V(\m^{\mu}_U-\m^{\mu}_V)(\m^{\mu}_U-\m^{\mu}_V)^T,
\end{array}
$$
where $p_U=\frac{\mu(U)}{\mu(U)+
\mu(V)},\, p_V:=\frac{\mu(V)}{\mu(U)+\mu(V)}$.

b) Assume that $V \subset U$ is such that $\mu(V)<\mu(U)$.  Then
$$
\begin{array}{l}
\m^{\mu}_{U \setminus V} =  q_U \m^{\mu}_U-q_V\m^{\mu}_V, \\[1ex]
\Sigma^{\mu}_{U \setminus V} = 
%\\ \phantom{=}
q_U\Sigma^{\mu}_U -q_V\Sigma^{\mu}_{V}
-q_Uq_V(\m^{\mu}_U-\m^{\mu}_V)(\m^{\mu}_U-\m^{\mu}_V)^T,
\end{array}
$$
where $q_U:=\frac{\mu(U)}{\mu(U)-\mu(V)},\, q_V:=\frac{\mu(V)}{\mu(U)-\mu(V)}$.
\end{theorem}

%% file: section/cluster.tex
\section{Clustering with respect to Gaussian families}

%%%%%%%%%%%%%%%%%%%%%%%%%%%%%%%%%%
\subsection{Introduction to clustering}

In the proceeding part of our paper we study the applications of our theory for clustering, where by {\em clustering
we understand division of the data into groups of similar type.}
Therefore since in clustering we consider only one fixed
subdensity family $\F$ we will use the notation
\begin{equation} \label{bbas}
h_{\mu}(\F;(U_i)_{i=1}^k):=\sum_{i=1}^k h_{\mu}(\F;U_i),
\end{equation}
for the family $(U_i)_{i=1}^k$ of pairwise disjoint Lebesgue measurable sets.  We see that \eqref{bbas} gives the total memory cost of disjoint $\F$-clustering of $(U_i)_{i=1}^k$.

The aim of $\F$-clustering is to find a $\mu$-partition $(U_i)_{i=1}^k$ (with possibly empty elements) which minimizes
$$
\begin{array}{l}
\ce{\oti \limits_{i=1}^k (\F|U_i)}=h_{\mu}(\F;(U_i)_{i=1}^k)
=\sum \limits_{i=1}^k \mu(U_i) \cdot [-\ln (\mu(U_i))+H^{\times}(\mu_{U_i}\|\F)].
\end{array}
$$
Observe that the amount of sets $(U_i)$ with nonzero $\mu$-measure
gives us the number of clusters into which we have divided our space.

In many cases we want the clustering to be independent
of translations, change of scale, isometry, etc.

\begin{definition}
Suppose that we are given a probability measure $\mu$.
We say that the clustering is {\em $A$-invariant} if instead of clustering $\mu$
we will obtain the same effect by
\begin{itemize}
\item introducing $\mu_A:=\mu \circ A^{-1}$ (observe that if $\mu$
corresponds to the data $\y$ then $\mu_A$ corresponds to the set $A(\y)$);
\item obtaining the clustering $(V_i)_{i=1}^k$ of $\mu_A$;
\item taking as the clustering of $\mu$ the sets $U_i=A^{-1}(V_i)$.   
\end{itemize}
\end{definition}

This problem is addressed in following observation which is 
a direct consequence of Corollary \ref{obs:aff}:

\begin{observation} \label{4.1}
Let $\F$ be a given subdensity family and $A$ be an affine 
invertible map. Then
$$
\begin{array}{l}
H^{\times}\big(\mu\|\oti \limits_{i=1}^k(\F|U_i)\big)
=H^{\times}\big(\mu \circ A^{-1}\|\oti \limits_{i=1}^k(\F_A|A(U_i))\big)+
\ln|\det A|.
\end{array}
$$
\end{observation}

As a consequence we obtain that if $\F$ is $A$-invariant, that is 
$\F=\F_A$, then the $\F$ clustering is also $A$-invariant.

The next important problem in clustering theory is the question how
to verify cluster validity. Cross entropy theory
gives a simple and reasonable answer -- namely from
the information point of view the clustering 
$$
U=U_1 \cup \ldots \cup U_k
$$ 
is profitable if we gain on separate compression by division into $(U_i)_{i=1}^k$, that is when:
$$
h_\mu(\F;(U_i)_{i=1}^k) < h_\mu(\F;U).
$$
This leads us to the definition of {\em $\F$-divergence of 
the splitting $U=U_1 \sqcup \ldots \sqcup U_k$}:
$$
\begin{array}{l}
d_\mu(\F;(U_i)_{i=1}^k):=h_\mu(\F;U)-h_\mu(\F;(U_i)_{i=1}^k).
\end{array}
$$
Trivially if $d_\mu(\F;(U_i)_{i=1}^k) > 0$ then we gain in using clusters $(U_i)_{i=1}^k$. Moreover, if $(U_i)_{i=1}^k$ is
a $\mu$-partition then
$$
d_\mu(\F;(U_i)_{i=1}^k)=\ce{\F}-\ce{\oti_{i=1}^k(\F|U_i)}.
$$
Observe that the above formula is somewhat reminiscent of the classical Kullback-Leibler divergence. 

%%%%%%%%%%%%%%%%%%%%%%%%%%%%%%%%%%
\subsection{Gaussian Clustering}

There are two most important clustering families one usually considers,
namely subfamilies of gaussian densities and of uniform densities.
In general gaussian densities are easier to use, faster in implementations, and more often appear in ``real-life'' data. However, in some cases the use of uniform densities is preferable
as it gives strict estimations for the belonging of the data points\footnote{For
example in computer games we often use bounding boxes/elipsoids
to avoid unnecessary verification of non-existing collisions.}. 

\begin{remark}
Clearly from uniform families, due to their affine invariance and ``good'' covering properties, most important are uniform densities on ellipsoids. Let us mention that the clustering of a dataset $\y$ by ellipsoids described
in \cite{ellips1} which aims to find the partition $\y=\y_1 \cup \ldots \cup
\y_k$ which minimizes
$$
-\sum_{i=1}^k w_i \cdot \ln \det(\Sigma_{\y_i}),
$$
where $w_i$ are weights, is close to CEC based on uniform densities on ellipsoids in $\R^N$ which, as one can easily check, reduces to the minimization of:
$$
- \sum_{i=1}^k p(\y_i) \cdot \ln \det (\Sigma_{\y_i})-\frac{2}{N}\sum_{i=1}^k \sh(p(\y_i)),
$$
where $p(\y_i):=\card(\y_i)/\card(\y)$.
\end{remark}

From now on we fix our attention on Gaussian clustering
(we use this name instead $\G$-clustering).
By Observation \ref{4.1} we obtain that the Gaussian clustering is invariant with 
respect to affine transformations.

By joining Proposition \ref{cross-gaussians} with \eqref{bbas} we obtain the basic formula on the Gaussian cross-entropy.

\begin{observation}
Let $(U_i)_{i=1}^k$ be a sequence of pairwise disjoint measurable sets. Then
\begin{equation} \label{basgauss2}
\begin{array}{l}
h_\mu(\G;(U_i)_{i=1}^k) =
\sum \limits_{i=1}^k \mu(U_i) \! \cdot \![\frac{N}{2}\ln(2\pi e)-\ln(\mu(U_i)) 
+\frac{1}{2} \ln \det(\Sigma_{U_i}^{\mu})].
\end{array}
\end{equation}
\end{observation}

In the case of Gaussian clustering due to the large
degree of freedom we were not able to obtain in the 
general case a simple formula 
for the divergence of two clusters. However,
we can easily consider the case of two
groups with equal covariances.

\begin{theorem}
Let us consider disjoint sets $U_1,U_2 \subset \R^N$ with identical 
covariance matrices $\Sigma^\mu_{U_1}=\Sigma^\mu_{U_2}=\Sigma$.
Then
$$
\begin{array}{l} 
d_\mu(\G;(U_1,U_2))/(\mu(U_1)+\mu(U_2))
=\frac{1}{2}\ln(1+p_1p_2 \|\m_{U_1}^\mu-\m_{U_2}^\mu\|_{\Sigma}^2)
-\sh(p_1)-\sh(p_2),
\end{array}
$$
where $p_i=\mu(U_i)/(\mu(U_1)+\mu(U_2))$.

Consequently $d_\mu(\G;(U_1,U_2)) >0$ iff
\begin{equation} \label{niez}
\|\m_{U_1}^\mu-\m_{U_2}^\mu\|_\Sigma^2 >p_1^{-2p_1-1}p_2^{-2p_2-1}-p_1^{-1}p_2^{-1}.
\end{equation}
\end{theorem}

\begin{proof}
By \eqref{basgauss2}
$$
\begin{array}{l}
d_\mu(\G;(U_1,U_2))/(\mu(U_1)+\mu(U_2))= 
\frac{1}{2}[\ln \det(\Sigma^{\mu}_{U_1 \cup U_2})-\ln \det (\Sigma) ]
-\sh(p_1)-\sh(p_2).
\end{array}
$$
By applying Theorem \ref{lop} the 
value of $\Sigma^\mu_{U_1 \cup U_2}$ simplifies to
$\Sigma+p_1p_2\m\m^T$, where $\m=(\m^\mu_{U_1}-\m^\mu_{U_2})$, 
and therefore we get
$$
\begin{array}{l}
d_\mu(\G;(U_1,U_2))/(\mu(U_1)+\mu(U_2)) \\
= \frac{1}{2}\ln \det(\I+p_1p_2\Sigma^{-1/2}\m\m^T\Sigma^{-1/2})-\sh(p_1)-\sh(p_2) \\[0.5ex]
= \frac{1}{2}\ln \det (\I+p_1p_2(\Sigma^{-1/2}\m)(\Sigma^{-1/2}\m)^T)-\sh(p_1)-\sh(p_2).
\end{array}
$$
Since $\det(I+\alpha v v^T)=1+\alpha \|v\|^2$ (to see this it suffices
to consider the matrix of the operator $I+\alpha v v^T$ in the orthonormal base which first element is $v/\|v\|$), we arrive at
$$
\begin{array}{l} 
d_\mu(\G;(U_1,U_2))/(\mu(U_1)+\mu(U_2))
=\frac{1}{2}\ln(1+p_1p_2 \|\m\|_{\Sigma}^2)-\sh(p_1)-\sh(p_2).
\end{array}
$$
Consequently $d_\mu(\G;(U_1,U_2)) >0$ iff
$$
\ln(1+p_1p_2\|\m\|_{\Sigma}^2)>2\sh(p_1)+2\sh(p_2),
$$
which is equivalent to 
$
1+p_1p_2\|\m\|_\Sigma^2 >p_1^{-2p_1}p_2^{-2p_2}
$.
\end{proof}

\begin{remark}
As a consequence of \eqref{niez} we obtain that
if the means of $U_1$ and $U_2$ are sufficiently
close in the Mahalanobis $\|\cdot\|_{\Sigma}$ distance, then it
is profitable to glue those sets together into one cluster.

Observe also that the constant in RHS of \eqref{niez}
is independent of the dimension. We mention it as an analogue
does not hold for Spherical clustering, see Observation \ref{sfer}.
\end{remark}

\begin{example}
\begin{figure}[!t]
\centering
	\subfigure[Plot of Gaussian divergence.]{\label{fig:plot}
\fbox{\includegraphics[height=0.2\textheight]{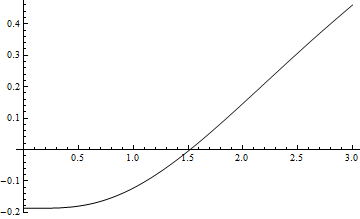}}}
	\subfigure[Densities.]{\label{fig:dens}
\fbox{\includegraphics[height=0.2\textheight]{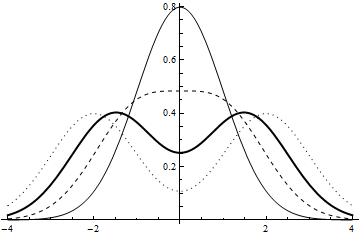}}}
\caption{Convex combination of gaussian densities. Black thick
density is the bordering density between one and two clusters} 
\label{fig:kull} 
\end{figure}
Consider the probability measure $\mu_s$ on $\R$ given as the convex combination of two gaussians with means at $s$ and $-s$, with density
$$
f_{s}:=\mbox{$\frac{1}{2}$}\Nor(s,1)+\mbox{$\frac{1}{2}$}\Nor(-s,1),
$$
where $s\geq 0$. Observe that with $s\to \infty$ the initial density $\Nor(0,1)$ separates into two almost independent gaussians.

To check for which $s$ the Gaussian divergence will see this behavior, we fix the partition $(-\infty,0),(0,\infty)$. One can easily verify that
$$
\begin{array}{l}
d_{\mu_s}\big(\G;((-\infty,0),(0,\infty))\big)=\\
-\ln(2)+\frac{1}{2} \ln(1+s^2) -\frac{1}{2} \ln[1-\frac{2 e^{-s^2}}{\pi }
%\\\phantom{-\frac{1}{2} \ln\big(}
+s^2\!-\sqrt{\frac{8}{\pi }}s e^{-\frac{s^2}{2}} \mathrm{Erf}(\frac{s}{\sqrt{2}})-s^2 \mathrm{Erf}(\frac{s}{\sqrt{2}})^2].
\end{array}
$$
Consequently, see Figure \ref{fig:plot}, there exists $s_0 \approx
1.518$ such that the clustering of $\R$ into two clusters $((-\infty,0),(0,\infty))$ is profitable iff $s >s_0$. On figure \ref{fig:dens} we show densities $f_s$ for $s=0$ (thin line); $s=1$ (dashed line); $s=s_0$ (thick line) and $s=2$ (points).

This theoretical result which puts the border between one
and two clusters at $s_0$ seems consistent with our geometrical intuition
of clustering of $\mu_s$.
\end{example}

%%%%%%%%%%%%%%%%%%%%%%%%
\subsection{Spherical Clustering}

In this section we consider spherical clustering which can be seen
as a simpler version of the Gaussian clustering.
By Observation \ref{4.1} we obtain that Spherical clustering is invariant with 
respect to scaling and isometric transformations (however, it is obviously not
invariant with respect to affine transformations).

\begin{observation}
Let $(U_i)_{i=1}^k$ be a $\mu$-partition. Then
\begin{equation} \label{43}
\begin{array}{l}
h_\mu(\S;(U_i)_{i=1}^k) =
\sum \limits_{i=1}^k \mu(U_i) \!\cdot\! [\frac{N}{2}\ln(2\pi e/N)
-\ln(\mu(U_i))+\frac{N}{2} \ln D_{U_i}^{\mu}].
\end{array}
\end{equation}
\end{observation}

To implement Hartigan approach to Spherical CEC and to 
deal with Spherical divergence the following
trivial consequence of Theorem \ref{lop} is useful.

\begin{corollary} \label{basi}
Let $U,V$ be measurable sets. 

a) Assume additionally that $U \cap V=\emptyset$ and $\mu(U)>0,\mu(V)>0$. Then 
$$
\begin{array}{rcl}
\m^{\mu}_{U \cup V} & = & p_{U}\m^{\mu}_{U}+p_V\m^{\mu}_V, \\[1ex]
D^{\mu}_{U \cup V} & = & p_UD^{\mu}_U+p_VD^{\mu}_V
+p_Up_V\|\m^{\mu}_U-\m^{\mu}_V\|^2,
\end{array}
$$
where $p_U:=\frac{\mu(U)}{\mu(U)+
\mu(V)},\, p_V:=\frac{\mu(V)}{\mu(U)+\mu(V)}$.

b) Assume that $V \subset U$ is such that $\mu(V)<\mu(U)$.  Then
$$
\begin{array}{rcl}
\m^{\mu}_{U \setminus V} & = & q_U \m^{\mu}_U-q_V\m^{\mu}_V, \\[1ex]
D_{U \setminus V} & = &
q_UD^{\mu}_U -q_VD^{\mu}_{V}
-q_Uq_V\|\m^{\mu}_U-\m^{\mu}_V\|^2,
\end{array}
$$
where $q_U:=\frac{\mu(U)}{\mu(U)-\mu(V)},\, q_V:=\frac{\mu(V)}{\mu(U)-\mu(V)}$.
\end{corollary}

\begin{example}
\begin{figure}[!t]
\centering
%	\subfigure[Mouse-like set.]{\label{fig:m_1}\fbox{\includegraphics[width=1.1in]{figure_5/mouse}}}
%	\subfigure[k-means with $k=3$.]{\label{fig:m_2}\fbox{\includegraphics[width=1.1in]{figure_5/mouse_k-means}}}
%     \subfigure[Spherical CEC.]{\label{fig:m_3}\fbox{\includegraphics[width=1.1in]{figure_5/mouse-sfer}}}	\\
	\subfigure[Three circles.]{\label{fig:n_1}\fbox{\includegraphics[width=1.2in]{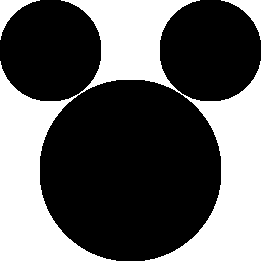}}}	\subfigure[k-means with $k=3$.]{\label{fig:n_2}\fbox{\includegraphics[width=1.2in]{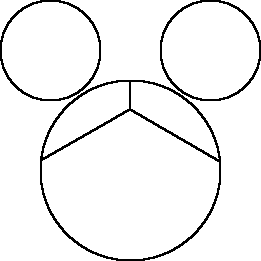}}}
     \subfigure[Spherical CEC.]{\label{fig:n_3}\fbox{\includegraphics[width=1.2in]{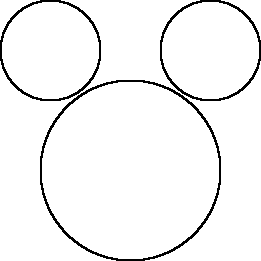}}}		
\caption{Comparison of classical k-means clustering (with $k=3$ clusters) with Spherical CEC (with initially $10$ clusters) of three ,,mouse-like'' %set and of three
circles.}
	\label{fig:mm} 
\end{figure}
We considered the uniform distribution on 
%mouse-like set Figure \ref{fig:m_1}
%and on 
the set consisting of three disjoint circles%\footnote{More precisely we consider the uniform density on the approximation of those sets build from many small squares} Figure \ref{fig:n_1}. 
We started CEC with initial choice of 10 clusters, as a result of Spherical 
CEC we obtained %in both cases 
clustering into three circles
%sets, 
see Figure
%\ref{fig:m_3} and 
\ref{fig:n_3} -- compare this result with the classical
k-means with $k=3$ on Figure
% \ref{fig:m_2} and 
\ref{fig:n_2}. %Observer that in
Observe that contrary
to classical k-means in spherical clustering we do not obtain the ``mouse effect''.
\end{example}

Let us now consider when we should join two groups. 

\begin{theorem}
Let $U_1$ and $U_2$ be disjoint measurable sets with nonzero $\mu$-measure.
We put $p_i=\mu(U_i)/(\mu(U_1)+\mu(U_2))$ and
$\m_i=\m^{\mu}_{U_i}$, $D_i=D^{\mu}_{U_i}$ for $i=1,2$.
Then
$$
\begin{array}{l} 
d_\mu(\G;(U_1,U_2))/(\mu(U_1)+\mu(U_2))\\[0.5ex]
=\frac{N}{2}\ln(p_1D_1+p_2D_2+p_1p_2\|\m_1-\m_2\|^2) 
%\\\phantom{=}
-p_1\frac{N}{2}\ln D_1-p_2\frac{N}{2}\ln D_2
-\sh(p_1)-\sh(p_2).
\end{array}
$$

Consequently,
$
d_\mu(\S;(U_1,U_2)) > 0
$
iff
$$
\|\m_1-\m_2\|^2 > \frac{D_1^{p_1}D_2^{p_2}}{p_1^{2p_1/N}p_2^{2p_2/N}}-(p_1D_1+p_2D_2).
$$
\end{theorem}

\begin{proof}
By \eqref{43}
$$
\begin{array}{l}
d_\mu(\S;(U_1,U_2))/(\mu(U_1)+\mu(U_2))\\[1ex]
=\frac{N}{2}\ln(D^\mu_{U_1 \cup U_2}) -p_1\frac{N}{2}\ln D_1
-p_2\frac{N}{2}\ln D_2-\sh(p_1)-\sh(p_2).
\end{array}
$$
Since by Corollary \ref{basi} 
$$
D^\mu_{U_1+U_2}=p_1D_1+p_2D_2+p_1p_2\|\m_1-\m_2\|^2,
$$ 
we obtain that $d_\mu(\S;(U_1,U_2))>0$ iff
$$
\|\m_1-\m_2\|^2 > \frac{D_1^{p_1}D_2^{p_2}}{p_1^{2p_1/N}p_2^{2p_2/N}}- 
(p_1D_1+p_2D_2).
$$
\end{proof}

\begin{observation} \label{sfer}
Let us simplify the above formula in the case when we have sets with identical measures $\mu(U_1)=\mu(U_2)$ and $D:=D^\mu_{U_1}=D^{\mu}_{U_2}$. Then by the previous theorem we should glue the groups together if
$$
\|\m_1-\m_2\| \leq \sqrt{4^{1/N}-1} \cdot r,
$$
where $r=\sqrt{D}$. So, as we expected, when the distance between the groups is proportional to their ``radius'' the joining becomes profitable.

Another, maybe less obvious, consequence of 
$$
4^{1/N}-1 \approx \frac{\ln 4}{N} \to 0 \mbox{ as } N \to \infty
$$
is that with the dimension $N$ growing we should join the groups/sets together if their centers become closer. This follows from the 
observation that if we choose two balls in $\R^N$ with radius $r$ and distance 
between centers $R\geq 2r$, the proportion of their volumes to the volume
of the containing ball decreases to zero with dimension
growing to infinity. 
\end{observation}

%%%%%%%%%%%%%%%%%%%%%%%%%%%
\subsection{Fixed covariance}

In this section we are going to discuss the simple case when
we cluster by $\G_{\Sigma}$, for a fixed $\Sigma$.
By Observation \ref{4.1} we obtain that $\G_{\Sigma}$ clustering is 
translation invariant (however, it is obviously not
invariant with respect to scaling or isometric transformations).

\begin{observation}
Let $\Sigma$ be fixed positive symmetric matrix.
and let $(U_i)_{i=1}^k$ be a sequence of pairwise disjoint
measurable sets. Then
$$
\begin{array}{l}
h_\mu(\G_{\Sigma};(U_i)_{i=1}^k) \\[1ex]
=\sum \limits_{i=1}^k \mu(U_i) \! \cdot \! (\frac{N}{2} \ln(2\pi)+\frac{1}{2}
\ln \det(\Sigma)) +
\sum \limits_{i=1}^k \mu(U_i)\! \cdot \! [-\ln(\mu(U_i))+\frac{1}{2}\tr(\Sigma^{-1}\Sigma^{\mu}_{U_i})].
\end{array}
$$
\end{observation}

This implies that in the $\G_{\Sigma}$ clustering we search
for the partition $(U_i)_{i=1}^k$ which minimizes
$$
\sum \limits_{i=1}^k \mu(U_i) \cdot \big[-\ln(\mu(U_i))+\frac{1}{2}\tr(\Sigma^{-1}\Sigma^{\mu}_{U_i})\big].
$$

Now we show that in the $\G_{\Sigma}$ clustering, if we have two groups with centers/means sufficiently close, it always pays to ``glue'' the groups together
into one.

\begin{theorem}
Let $U_1$ and $U_2$ be disjoint measurable sets with nonzero $\mu$-measure.
We put $p_i=\mu(U_i)/(\mu(U_1)+\mu(U_2))$.
Then
\begin{equation}\label{prr}
\begin{array}{l} 
d_\mu(\G_\Sigma;(U_1,U_2))/(\mu(U_1)+\mu(U_2))%\\[0.5ex]
=p_1p_2\|\m^\mu_{U_1}-\m^\mu_{U_2}\|_\Sigma^2-\sh(p_1)-\sh(p_2).
\end{array}
\end{equation}
Consequently
$
d_\mu(\G_\Sigma;(U_1,U_2)) > 0
$
iff
$$
\|\m^\mu_{U_1}-\m^\mu_{U_2}\|_{\Sigma}^2 > \frac{\sh(p_1)+\sh(p_2)}{p_1p_2}.
$$
\end{theorem}

\begin{proof}
We have
\begin{equation}\label{simp}
\begin{array}{l}
d_\mu(\G_\Sigma;(U_1,U_2))/(\mu(U_1)+\mu(U_2))\\[1ex]
=\frac{1}{2}\tr(\Sigma^{-1}\Sigma^\mu_{U_1 \cup U_2})%\\[0.5ex]
-\frac{p_1}{2}\tr(\Sigma^{-1}\Sigma^\mu_{U_1})-\frac{p_2}{2}\tr (\Sigma^{-1}\Sigma^\mu_{U_2})-\sh(p_1)-\sh(p_2).
\end{array}
\end{equation}
Let $\m=\m^{\mu}_{U_1}-\m^{\mu}_{U_2}$. Since
$
\Sigma^{\mu}_{U_1 \cup U_2} = 
p_1\Sigma^{\mu}_{U_1}+p_2\Sigma^{\mu}_{U_2}
+p_1p_2\m\m^T,
$
and $\tr(AB)=\tr(BA)$, \eqref{simp} simplifies to \eqref{prr}.
\end{proof}

Observe that the above formula is independent of 
deviations in groups, but only on the distance of the centers
of weights (means in each groups).

\begin{lemma}
The function 
$$
\{(p_1,p_2) \in (0,1)^2: p_1+p_2=1\} \to \frac{\sh(p_1)+\sh(p_2)}{p_1p_2}
$$
attains global minimum $\ln 16$ at $p_1=p_2=1/2$.
\end{lemma}

\begin{proof}
Consider 
$$
w:(0,1) \ni p \to \frac{\sh(p)+\sh(1-p)}{p(1-p)}. 
$$
Since $w$ is symmetric with respect to $1/2$, to show assertion it is sufficient to prove that $w$ is convex.

We have
$$
\!w''(p)\!=\!\frac{2 (-1\!+\!p)^3 \ln(1\!-\!p)+p \left(-1\!+\!p\!-\!2 p^2 \ln(p)\right)}{(1\!-\!p)^3 p^3}.
$$
Since the denominator of $w''$ is nonnegative, we consider only
the numerator, which we denote by $g(p)$.
The fourth derivative of $g$ equals $12/[p(1-p)]$.
This implies that
$$
g''(p)=4 (-2+3 (-1+p) \ln(1-p)-3 p \ln(p))
$$
is convex, and since it is symmetric around $1/2$, it has the global
minimum at $1/2$ which equals
$$
g''(1/2)=4(-2+3\ln 2)=4\ln(8/e^2)>0.
$$
Consequently $g''(p)>0$ for $p \in (0,1)$, which implies
that $g$ is convex. Being symmetric around $1/2$ it attains
minimum at $1/2$ which equals
$
g(1/2)=\frac{1}{4}\ln(4/e)>0,
$
which implies that $g$ is nonnegative, and consequently $w''$ is
also nonegative. Therefore $w$ is convex and symmetric around $1/2$,
and therefore attains its global minimum $4 \ln 2$ at $p=1/2$.
\end{proof}

\begin{corollary} \label{poi}
If we have two clusters with centers $\m_1$ and $\m_2$, then it is always
profitable to glue them together into one group in $\G_\Sigma$-clustering if
$$
\|m_1-m_2\|_{\Sigma} < \sqrt{\ln 16} \approx 1.665.
$$
\end{corollary}

As a direct consequence we get:

\begin{corollary}
Let $\mu$ be a measure with support contained in a bounded convex
set $V$. Then the number of clusters which realize the cross-entropy $\G_\Sigma$ is bounded from above by the maximal cardinality of an $\e$-net
(with respect to the Mahalanobis distance $\|\cdot\|_{\Sigma}$), where $\e=\sqrt{4 \ln2}$, in $V$.
\end{corollary}

\begin{proof}
By $k$ we denote the maximal cardinality of the $\e$-net with respect
to the Mahalanobis distance.

Consider an arbitrary $\mu$-partition $(U_i)_{i=1}^l$ consisting
of sets with nonempty $\mu$-measure. Suppose
that $l>k$. We are going to construct a $\mu$-partition with $l-1$ elements
which has smaller cross-entropy then $(U_i)$.

To do so consider the set $(\m^\mu_{U_i})_{i=1}^l$ consisting of centers of 
the sets $U_i$. By the assumptions we know that there exist at least two 
centers which are closer then $\e$ -- for simplicity assume that
$\|\m^\mu_{U_{l-1}}-\m^\mu_{U_l}\|_\Sigma <\e$. Then by the previous results we obtain that
$$
h_\mu(\G_\Sigma;U_{l-1} \cup U_l) < h_\mu(\G_\Sigma;U_{l-1})+h_\mu(\G_\Sigma;U_l).
$$
This implies that the $\mu$-partition $(U_1,\ldots,U_{l-2},U_{l-1}
\cup U_l)$ has smaller cross-entropy then $(U_i)_{i=1}^l$.
\end{proof}

%%%%%%%%%%%%%%%%%%%%%%%%%%
\subsection{Spherical CEC with scale and $k$-means}

We recall that $\G_{\s\I}$ denotes the set of all normal densities with covariance $s\I$.
We are going to show that for $s \to 0$ results of $\G_{s\I}$-CEC converge to k-means clustering, while for $s \to \infty$ our data will form one big group.

\begin{observation}
For the sequence $(U_i)_{i=1}^k$ we get
$$
\begin{array}{l}
h_\mu(\G_{s\I};(U_i)_{i=1}^k) 
=\sum \limits_{i=1}^k \mu(U_i)\cdot (\frac{N}{2}\ln(2\pi s)-\ln \mu(U_i)+\frac{N}{2s}D^{\mu}_{U_i}).
\end{array}
$$
\end{observation}

Clearly by Observation \ref{4.1} $\G_{s\I}$ clustering is isometry invariant, 
however it is not scale invariant.  

\medskip

To compare k-means with Spherical CEC with fixed scale
let us first describe classical k-means from our point of view.
Let $\mu$ denote the discrete or continuous probability measure.
For a $\mu$-partition $(U_i)_{i=1}^k$ we introduce the
{\em within clusters sum of squares} by the formula
$$
\begin{array}{l}
\wcss(\mu\|(U_i)_{i=1}^k):=\sum \limits_{i=1}^k
\int_{U_i}\|x-\m^\mu_{U_i}\|^2d\mu(x) \\[1ex]
=\sum \limits_{i=1}^k
\mu(U_i)\int \!\|x-\m^\mu_{U_i}\|^2d\mu_{U_i}(x)
=\sum \limits_{i=1}^k \mu(U_i) \!\cdot\! D_{U_i}^\mu.
\end{array}
$$
\begin{remark}
Observe that if we have data $\y$ partitioned into $\y=\y_1 \cup
\ldots \cup \y_k$, then the above coincides
(modulo multiplication by the cardinality of $\y$) with the classical within clusters sum of squares. Namely, for discrete probability measure 
$
\mu_Y:=\frac{1}{\card(\y)}\sum \limits_{y \in \y}\d_y
$
we have
$
\wcss(\mu_\y\|(\y_i)_{i=1}^k)=
\frac{1}{\card(\y)} \sum\limits_{i=1}^k \sum\limits_{y \in \y_i} \|y-\m_{\y_i}\|^2
$.
\end{remark}

In classical k-means
the aim is to find such $\mu$-partition $(U_i)_{i=1}^k$ which 
minimizes the within clusters sum of squares
\begin{equation} \label{kmeans}
%S(\mu\|(U_i)_{i=1}^k):=
\sum_{i=1}^k \mu(U_i) \cdot D^{\mu}_{U_i},
\end{equation}
while in $\G_{s\I}$-clustering our aim is to minimize
$$
\sum_{i=1}^k \mu(U_i)\cdot (-\frac{2s}{N}\ln \mu(U_i)+D^{\mu}_{U_i}).
$$
Obviously with $s \to 0$, the above function converges to
\eqref{kmeans}, which implies that k-means
clustering can be understood as the limiting case of $\G_{s\I}$ clustering,
with $s \to 0$. 

\begin{example}
\begin{figure}[!t]
\centering
	\subfigure[k-means with $k\!=\!5$.\!]{\label{fig:sqr_2_a}\fbox{\includegraphics[width=1.5in]{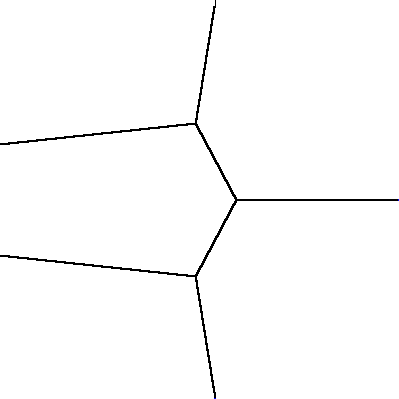}}}
	\subfigure[$\G_{s\I}$-CEC for $s\!=\!5\!\cdot 10^{-5}$ and $5$ clusters.\!]{\label{fig:sqr_2_a}\fbox{\includegraphics[width=1.5in]{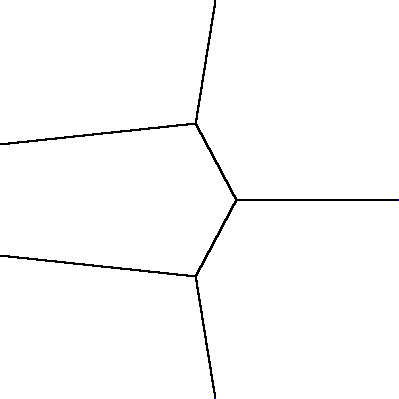}}}
	\label{fig:voris} 
\end{figure}
We compare on Figure \ref{fig:voris} $\G_{s\I}$ clustering of the square $[0,1]^2$ with very small $s=5 \cdot 10^{-5}$ to k-means. As we see we obtain optically identical results.
\end{example}

\begin{observation} \label{obs-kmeans}
We have
$$
\begin{array}{l}
0 \leq \wcss(\mu\|(U_i)_{i=1}^k) -[-s\ln(2\pi s)+\frac{2s}{N}
\ce{\oti \limits_{i=1}^k (\G_{s\I}|U_i)}]\\
=\frac{s}{2N}\sum \limits_{i=1}^k
\mu(U_i) \cdot \ln(\mu(U_i)) \leq \frac{\ln (k)}{2N}s.
\end{array}
$$
This means that for an arbitrary partition consisting of k-sets $\wcss(\mu\|\cdot)$ can 
be approximated (as $s \to 0$) with the affine combination of
$\ce{\G_{s\I}}$, which can be symbolically summarized as interpretation of k-means as $\G_{0 \cdot \I}$ clustering.
\end{observation}

If we cluster with $s \to \infty$ we have
tendency to build larger and larger clusters. 

\begin{proposition}
Let $\mu$ be a measure with support of diameter $d$.
Then for
$$
s>\frac{d^2}{\ln 16}
$$ 
the optimal clustering with respect to $\G_{s\I}$ will be obtained for one large group.

More precisely, for every $k>1$ and $\mu$-partition $(U_i)_{i=1}^k$ 
consisting of sets of nonempty $\mu$-measure we have
$$
\ce{\F} < \ce{\oti_{i=1}^k (\F|U_i)}.
$$
\end{proposition}

\begin{proof}
By applying Corollary \ref{poi} with $\Sigma=s\I$ we obtain that we should 
always glue two groups with centers $\m_1,\m_2$ together if
$\|\m_1-\m_2\|^2_{s\I} < \ln 16$,  or equivalently if 
$
\|\m_1-\m_2\|^2 < s\ln 16
$.
\end{proof}

Concluding, if the radius tends to zero, we cluster the data into smaller and
smaller groups, while for the radius going to $\infty$, the data will have the tendency to form only one group.